\documentclass[%
reprint,
amsmath,amssymb,
aps,
prb,
]{revtex4-2}

\usepackage{graphicx}% Include figure files
\usepackage{dcolumn}% Align table columns on decimal point
\usepackage{bm}% bold math
\usepackage{physics}
\usepackage{color,soul}
\usepackage{float}

\begin{document}

\newcommand{\En}[2]{ E^{\text{#1}}_{\text{#2}}\left[  \rho \right]}
\newcommand{\Ek}[3]{ E^{\text{#1}}_{\text{#2}}\left[  #3 \right]}
\newcommand{\Ep}[3]{  \epsilon^{\text{#1}}_{\text{#2}}\left(  #3 \right)}
\newcommand{\Fun}[3]{ F^{\text{#1}}_{\text{#2}}\left(  #3 \right)}
\newcommand{\den}{\rho\left(  \mathbf{r} \right)}

\newcommand{\Es}[2]{ E^{\text{#1}}_{\text{#2}}}

\title{Quasiparticle states of hexagonal BN: A van der Waals density functional study}% Force line breaks with \\

\author{Raul Quintero-Monsebaiz}
\email{raulq@chalmers.se}
\affiliation{%
 Microtechnology and Nanoscience, MC2, Chalmers University of Technology, SE-412 96 G$\Ddot{o}$teborg, Sweden  
}
\author{Per Hyldgaard}%
\email{hyldgaar@chalmers.se}
\affiliation{%
 Microtechnology and Nanoscience, MC2, Chalmers University of Technology, SE-412 96 G$\Ddot{o}$teborg, Sweden  
}%

\date{\today}

\begin{abstract}
We compute and track the impact of truly nonlocal-correlation effects on the quasi-particle (QP) band-structure of hexagonal boron-nitride (h-BN) systems. To that end, we start with the consistent-exchange vdW-DF-cx version [PRB \textbf{89}, 035412 (2014)] of the van der Waals  density functional 
(vdW-DF) method [JPCM \textbf{39}, 390001 (2020)] for exchange-correlation 
(XC) functional design and enforce piece-wise linearity in the energy changes with partial charging, using the Koopmans-integer (KI) DFT framework [JCTC \textbf{19}, 7079 (2023)]. Our approach and results (denoted KI-CX) extends present-standard use of KI DFT (denoted KI-PBE as it is based on the semilocal PBE [PRL \textbf{77}, 3865 (1996)] XC functional) to capture, for example, the impact of the interlayer coupling on the QPs.  We contrast KI-CX and KI-PBE results for the QP band-structure and compare with both \textit{GW} calculations and experimental observations of the (direct and indirect) QP gaps. We find that KI-CX brings improvements in the h-BN QP energy description and generally agrees with $GW$ studies.
\end{abstract}

\maketitle

\section{\label{sec:level1}Introduction\protect }

The goal of improving materials predictions by  parameter-free theory prompts us to seek deeper understanding of sudden charged excitations in light-matter interactions  
and of ground state (GS) properties in concert. For the former, we can summarize the nature of such excitations
as quasi-particle (QP) states, whose energies and spatial variation reflect the characteristics of the electronic-structure changes \cite{AuJoWi00,stephen2014,RanPRB16}
upon photon absorption or emission. Formally exact many body perturbation theory (MBPT), for example, in the GW$\Gamma$ \cite{hedin1965,reining2018}
or equation-of-motion
coupled-cluster \cite{stanton1994,nooijen1995,barlett1997,quintero2023}
(EOMCC) approaches, permit us to predict the QPs 
as solutions to effective dynamics problems. Meanwhile, for GS properties, formally exact density functional theory \cite{hohenberg1964} (DFT), using either the traditional and popular
Kohn-Sham (KS) \cite{kohn1965,gujo80,Seminario19951,burke}, the generalized-KS \cite{GKSstart,kronik2012,perdew1996,kronik2012,OTRSHalga,chen2018,skone2016,WiOhHa21,shukla_2022a,JPCM2025}, or the various Koopmans \cite{davo2010,ferretti2014,borghi2014,gennaro2022,colonna2022,linscott2023,koopweb2025} DFT frameworks, can predict the atomic structure and the electron-density variations. DFT also relies on an equation-of-motion framework but for fictitious independent 
particles in an effective local potential that reflects enough of the QP dynamics that we can compute the GS electron behavior \cite{larsstigssp,helujpc1971,shsl1983,casida1999,hyldgaard2020}. 

In practice, we need MBPT input to define robust so-called exchange-correlation (XC) functional approximations \cite{mabr,la70,langreth1975,gulu76,langreth1977,lape80,lameprl1981,lavo87,langreth1990,perdew21996,casida1999,thonhauser2007,hyldgaard2020}. Approximations also enter MBPT-based QP studies (e.g., in GW \cite{reining2018}) and we seek 
consistency in XC design work   \cite{helujpc1971,lape80,shsl1983,casida1999,chong2002,ferretti2014,hyldgaard2020,JPCM2025}. Progress in such a concerted approach should and can be tested on both GS and QP properties \cite{cococcioni2005,nguyen2018,WiOhHa21,JPCM2025}.

With an accurate XC functional approximation, KS-DFT lets us predict the atomic configurations \cite{burke} and materials vibrations, e.g., Refs.\  \cite{BrownAltvPRB16,grwahy20}.
Such computational characterizations permit theory validation both directly (on thermophysical properties and phases,
e.g., Ref.\ \cite{PeGrRo20}) and indirectly on the QPs, 
as revealed in electrical transport measurements \cite{yu2010}, scanning tunneling microscopy (STM) \cite{magonov1993}, angle-resolved photoemission (ARPES) \cite{zhang2022,grunert2020}, or inverse photoemission \cite{smith1988}.
These properties generally reflect 
the QP band structure (energy variation with momentum in an extended system). In terms of MBPT and XC quality testing, one key observation is that (sub-)picometer accuracy at atomic structure 
characterization is typically required to, in turn, accurately predict the light-matter interactions in the more costly MBPT methods \cite{RanPRB16}.

Notably, direct theory-experiment
testing on QPs has a higher value still for materials-theory development. We observe that Koopmans-integer (KI) DFT \cite{borghi2014,nguyen2018,gennaro2022,colonna2022,linscott2023}, using a linear-response description based on a given input XC functional, stands out by providing fast QP predictions
without changing neither total energies nor the forces that set structure when that XC input is used in KS-DFT \cite{linscott2023}. Accurate tools, linking the structure and QP aspects of materials modeling, are valuable for characterizations in their own rights but KI-DFT offers a potential for synenergy in continued method developments. The structure of the Hedin QP-equation-of-motion framework is equivalent to the formal adiabatic-connection formula (ACF)
specification of the 
XC functional designs 
\cite{helujpc1971,perdew21996,hyldgaard2020}. For a versatile XC functional, and certainly when used for layered materials, we expect to find fingerprints of, for example, van der Waals (vdW) interactions on the QPs band structure, that is, in the energy variation with the QP momentum. We can test, for example, the accuracy 
of an XC choice on descriptions of the interlayer vdW attractions by comparing the corresponding QP predictions by KI-DFT to either measurements or direct MBPT characterizations.

Present-standard KI-DFT studies \cite{Dabo2013,borghi2014,nguyen2015,nguyen2016,nguyen2018,colonna2018,colonna2019,Elliott2019,almeida2021,colonna2022,gennaro2022,linscott2023,marrazzo2024}, rely on either the local-density approximation (LDA) \cite{helujpc1971,pewa92} or the the semilocal 
Perdew-Burke-Ernzerhof
(PBE) \cite{perdew21996} XC functional. The latter of these functionals is developed for KS-DFT and gives what we here label as PBE results. Via an extension to density functional perturbation theory (DFPT), PBE also permits efficient QP characterizations in what we call KI-PBE, when
a piecewise-linearity (PWL) constraint \cite{perdew1982,davo2010,nguyen2018,linscott2023,schubert2023} is imposed on the energy variation with partial charging. However, KI-DFT still reflects the nature of the XC functional that is chosen for KI-DFT \cite{linscott2023} and KI-PBE is not set up to track vdW interactions \cite{hyldgaard2020}. This holds also for KI-PBE predictions of the QP band structure characterization, that is, in the QP energy variation with momentum.

It is motivated to move KI-DFT usage to instead start from a truly nonlocal-correlation functional form,
available within the 
vdW density functional (vdW-DF) method \cite{rydberg2003,dion2004,thonhauser2007,hybesc14,Berland_2015:van_waals,Thonhauser_2015:spin_signature,hyldgaard2020,shukla2022,JPCM2025}. Stacking multiple layers of a two-dimensional material to form a bulk system is known to alter its band structure \cite{wickramaratne2018}. An expansion that reflects an incorrect structure prediction of the interlayer spacing, will induce further changes \cite{hunt2020}.
Beyond those observations is again the fact that QPs are the most direct way that we can test the quality of the QP and MBPT logic behind many of our XC functionals designs \cite{helujpc1971,shsl1983,casida1999,perdew21996,rydberg2000,dion2004,hybesc14,berland2014,hyldgaard2020}.

Here we therefore present
a study tracking vdW signatures on QPs in hexagonal boron-nitride systems (h-BN) while introducing the use of KI-CX, that is, KI-DFT with a start in the consistent--exchange vdW-DF-cx version (here abbreviated CX)
\cite{berland2014,berland2014a,hyldgaard2020}.
We note that the \textsc{QuantumESPRESSO} (QE) suite of DFT codes \cite{giannozzi2009,giannozzi2017}, with the
module for \textsc{Koopmans}-DFT extension \cite{koopweb2025,linscott2023}, 
is already set up for KI-CX studies. The
h-BN systems are insulating and have been thoroughly studied both theoretically and experimentally. h-BN has emerged as one of the most prominent two-dimensional materials, owing in part to a graphene-like honeycomb structure in its layers and to its wide band gap \cite{caldwell2019,roy2021}. Additionally, h-BN exhibits a suite of advantageous properties, such as excellent thermal conductivity \cite{lindsay2011}, unique thickness-dependent behavior \cite{li2009}, chemical stability \cite{weng2016}, optical functionality \cite{zunger1976}, and polymer compatibility \cite{yu2018}. These attributes enable applications in aerospace \cite{zhang2017,weng2016}, environmental science, and energy technologies—spanning adsorption, photo-catalysis, membrane separation, disinfection, environmental sensing, as well as energy conversion and storage \cite{li2022}. 

The QPs correspond to sudden events that may be either adding or removing an electron \cite{AuJoWi00}. We focus on QPs in three systems that are experimentally available, and we first set the atomic structure by observations. This means that KI-CX versus KI-PBE differences 
reflect the direct impacts of truly nonlocal-correlation effects (interlayer coupling) 
on the QPs. We next track the full impact of including nonlocal-correlations (via KI-CX) on the QP description. This is done by also contrasting KI-CX and KI-PBE QP descriptions at the structure predictions provided by these XC functional choices (namely, using CX and PBE in KS-DFT, respectively).
There are local and global valence-band (conduction-band) extrema, together forming a number of (indirect) QP gaps in h-BN systems. We assess KI-PBE and KI-CX quality on accuracy for those gaps,
comparing with  experimental observations and GW studies. 

The manuscript is structured as follows. The following two sections summarizes the CX functional and use of KI-DFT, including computational parameters, input configurations, and work flows. Our results and discussion section presents and interprets KI-CX versus KI-PBE performance differences.  Section V contains our summary and outlook. The paper contains two appendices.

\section{Theory}

In this section, we first summarize the formulation of the non-local energy term $\Es{nl}{c}$ that enters in a vdW-DF like CX.  We next identify the components responsible for capturing non-local interactions in QP states when we use a CX starting point in KI-DFT.

\subsection{Truly nonlocal correlation in CX}

The design of the vdW-DFs relies on MBPT characterizations \cite{mabr,gulu76,langreth1977,lape80,lameprl1981,lavo87,langreth1990,thonhauser2007} of the density-density response function $\chi_{\lambda}\qty(\mathbf{r},\mathbf{r'},\omega)$, i.e., the ratio of the density change $\delta n(\mathbf{r})$ that arise when external potential at frequency $\omega$ is applied at position $\mathbf{r'}$. As indicated by the subscript, this response is determined
while assuming that the
raw inter-electron Coulomb
matrix element $V=|\mathbf{r}-\mathbf{r'}|^{-1}$ is scaled by a coupling constant $0 <  \lambda < 1$. The adiabatic connection formula (ACF)  \cite{langreth1975,gulu76,langreth1977,gunnarsson1977},
\begin{equation} 
E_{\text{xc}}[n] = -\int_{0}^{1} \! d\lambda \int_{0}^{\infty} \! \frac{du}{2\pi} \, \text{Tr}\left[ V \chi_{\lambda} \qty(iu) \right] - E_{\text{self}} \, , 
\label{eq:exc_acf}
\end{equation}
is used to convert the formal MBPT input into 
versions of an XC energy functional of the electron density variation $\rho(\mathbf{r})$.  The trace indicate integration with respect the spatial coordinates of $\chi_{\lambda}\qty(\mathbf{r},\mathbf{r'},\omega=iu)$ and $V$.

The resulting vdW-DF functionals are in practice given by treating the XC design work as an electrodynamics problem \cite{rydberg2000,rydberg2003,hybesc14}. Critical are the focus on collective excitations, compliance with Lindhard screening, and enforcing current conservation and 
time-reversal symmetry in modeling the local-field response \cite{dion2004,hybesc14,hyldgaard2020}. We start with a generalized-gradient approximation (GGA) for a
so-called internal XC function $E_\text{xc}^\text{in}[\rho]$,
comprising the XC terms of
the local-density approximation (LDA) \cite{helujpc1971,perdew1986}
and a MBPT-guided choice for the gradient-corrected exchange \cite{dion2004,lee10p081101,Berland_2015:van_waals,ChBeTh20}, This ensures a GGA-type exchange component $E_\text{x}^\text{in}[\rho]$ in the starting point for the vdW-DF design.
We next use an electrodynamics-based recast of the ACF \cite{rydberg2000,rydberg2003,hyldgaard2020} to extract a corresponding initial-guess
dielectric function 
$\epsilon(\mathbf{r},\mathbf{r'},\omega=iu)$ \cite{lee10p081101,Berland_2015:van_waals}:
\begin{equation} 
\Es{in}{xc} = \int \, \frac{du}{2\pi} \, \text{Tr}\left[ \text{ln} \qty(\epsilon \qty(iu)) \right] - \Es{}{self} \, .
\label{eq:Excindef} 
\end{equation}
This reformulation of the semilocal 
functional $\Es{in}{xc}$  reflects 
an exponential resummation, 
$\epsilon(\omega)\equiv 
\exp(S_\text{xc}(\omega))$,
of a 
plasmon-propagator $S_\text{xc}(\mathbf {r},\mathbf{r'},\omega) 
$  that corresponds to the
internal GGA \cite{Berland_2015:van_waals}, so that the initial-guess for the collective
excitations (found at  
$\epsilon(\omega)=0$) are
tracked by logarithmic singularities \cite{jerry65,hybesc14}. Third, we note that the Coulomb Green function $G$ is set up to trace also long-ranged electrodynamics couplings of virtual electron dynamics (for example, as reflected by plasmons) \cite{rydberg2003,hybesc14} and  that the vdW forces are
naturally included when we 
enforce current-conservation, 
$\epsilon(\omega) 
\longrightarrow \kappa(\omega) \equiv \nabla \epsilon(\omega) \cdot \nabla G$ \cite{rydberg2000,rydberg2003,dion2004,hybesc14}. We therefore use the resulting beyond-GGA
recast of the ACF for a formal vdW-DF-method specification \cite{lee10p081101,Berland_2015:van_waals,hyldgaard2020} 
\begin{equation} 
\Es{vdW-DF}{xc} = \int \, \frac{du}{2\pi} \, \text{Tr}\left[ \text{ln} \qty(\kappa \qty(iu)) \right] - \Es{}{self} \, .
\label{eq:EvdWDF} 
\end{equation}
Finally, we often (but not always \cite{rydberg2000,rydberg2003}) expand the truly nonlocal correlation term
\begin{equation}
E_{\text{c}}^{\text{nl}}[\rho] \equiv \Es{vdW-DF}{xc} -
\Es{in}{xc} \, ,
\label{eq:EcnlDEF} 
\end{equation}
in terms of both the Lindhard
screening an the exponential resummation to second order 
in $S_\text{xc}(\omega)$ \cite{dion2004,Berland_2015:van_waals,hyldgaard2020}:
\begin{equation} 
\Es{nl}{c} = \int \! \frac{du}{2\pi} \, \text{Tr} \left[ S_{xc}^2 - \qty(\nabla S_{xc} \cdot \nabla G )^2 \right]. 
\label{eq:Enl2} 
\end{equation}

The complete  
specification of the CX energy functional  \cite{berland2014,berland2014a} is denoted
$\Es{CX}{xc}$. It has an exchange component that mirrors that of $\Es{in}{xc}$ (as far as 
possible) to ensure XC balance in calculations of
energy differences \cite{berland2014,hyldgaard2020}. CX yields a local effective potential that defines the equation of motion for the KS-DFT orbitals. The consistent focus on the plasmon behavior means that the specification of the total correlation,  
$E_\text{c}^\text{LDA}
+E_{\text{c}}^{\text{nl}}$,
is explicitly given by the density. 
Using Eq.\ (\ref{eq:Enl2}) we can even 
cast the nonlocal-correlation energy functional \cite{dion2004,thonhauser2007}
 \begin{equation}
E_{\text{c}}^{\text{nl}}  = \frac{1}{2}\, \int \int \rho(\mathbf{r}) \Phi_c^\text{nl}[\rho](\mathbf{r},\mathbf{r'}) \rho(\mathbf{r'}) \, d\mathbf{r} \, d\mathbf{r'} \, ,
\label{eq:nlenergybykernel}
\end{equation}
with a spline representation \cite{roso09}
for what is a universal nonlocal-correlation kernel
$\Phi_c^\text{nl}[\rho]$.  This keeps computational costs comparable to that of any GGA-based KS-DFT \cite{roso09}, for example, PBE \cite{perdew21996}.

In a set of recent designs we have sought to design and leverage generalized-KS vdW-DFs \cite{levy1984,DFcx02017,cx0p2018,shukla_2022a,shukla2022,JPCM2025}. This is done in part to overcome residual accuracy issues in PBE or CX KS-DFT characterizations and 
sometimes also in the PBE-based range-separated hybrid (RSH), denoted HSE \cite{HSE03,HSE06,OTRSHalga,skone2016,chen2018,WiOhHa21}; The relevance of having access to global-hybrid RSH vdW-DFs is illustrated and summarized, 
for example, in Refs.\ 
\cite{PeGrRo20,grwahy20,frostenson2022,frostenson2024,JPCM2025}. Starting with two robust vdW-DF versions, CX and rev-vdW-DF2 \cite{hamada14}, and replacing, for example, part of the short-ranged GGA exchange, we arrive at two RSH vdW-DFs, denoted AHCX \cite{shukla_2022a} and AHBR \cite{shukla2022}. They generalize the HSE by providing a consistent inclusion 
of truly nonlocal-correlation effects 
and, in particular vdW interactions.
The AHCX and AHBR are general-purpose options, noting that they are applicable also to metallic systems \cite{shukla_2022a,shukla2022} and for a case when we expect strong static-correlation effects on molecular structure \cite{frostenson2024}. Still,
we can motivate adjustments of these basic RSH vdW-DF designs when used for molecular systems or for semiconductors: For the former (latter) we expect an absence (moderate presence) of asymptotic screening in the exchange descriptions, as implemented in our recent AHBR-mRSH \cite{JPCM2025} (in global-hybrid vdW-DFs like CX0p \cite{cx0p2018,grwahy20}). The logic is that the system-specific dielectric function is implicitly set by the Hartree-energy term and should therefore (ideally) also be reflected in the exchange description \cite{gulu76,kronik2012,OTRSHgap,skone2016,chen2018,WiOhHa21,JPCM2025}. It is hard to realize this goal in basic XC approximations that aims to be fast and universally applicable \cite{langreth1977,levy1984,perdew21996,berland2014}.

Importantly, practical use of such new hybrid vdW-DFs relies on the orbital-based evaluation of Fock exchange that are now available in QE \cite{giannozzi2017,linACE,PaoloElStruct1,shukla_2022a}. That is, our hybrid vdW-DF calculations, e.g.,  
\cite{DFcx02017,PeGrRo20,shukla2022,frostenson2022,frostenson2024}, automatically proceed in the generalized-KS framework \cite{levy1984}. This opens for vdW-inclusive QP predictions, when we apply also the ideas so-called optimal tuning \cite{kronik2012,OTRSHalga,OTRSHgap,OTRSHadsorp17,WiOhHa21} to our new molecule-focused AHBR-mRSH \cite{JPCM2025}. Excitingly, use of AHBR-mRSH with optimal tuning set by consistency, constitute a generalized KS-DFT that complies with the PWL constraint \cite{perdew1982,davo2010,OTRSHalga,WiOhHa21,JPCM2025} and has high accuracy for both binding energies and QPs
of nitrogen-base systems \cite{JPCM2025}.

The present focus on moving also KI-DFT to a vdW-DF start and our introduction of the
optimally-tuned AHBR-mRSH supplement each other. 
The same applies also (to some extent) for the more traditional RSH vdW-DF designs, AHCX and AHBR,
and when a  Hubbard-U term (U) \cite{liechtenstein1995, dudarev1998} is added 
to vdW-DF as the starting XC functional choice  \cite{pengperdew2017}. These methods restore PWL \cite{cococcioni2005} to varying extents and, for example, RSH vdW-DFs simultaneously improve parameter-free vdW-DF descriptions of total energies and structure \cite{shukla2022}. Also, AHBR-mRSH delivers QP predictions that employ the  same  nonlocal-correlation description \cite{lee10p081101}
that is already in wide KS-DFT usage \cite{hamada14,shukla2022,JPCM2025}. However, by including some Fock exchange in AHBR-mRSH we also simultaneously correct some density-driven errors, e.g, for transition states, that affects the underlying
rev-vdW-DF2 functional \cite{shukla2022,JPCM2025}.
In contrast, the here-introduced KI-CX approach (described below) leaves the GS energies and structure unchanged \cite{colonna2022,linscott2023} and therefore tests CX directly in terms of both GS and QP properties.

\subsection{CX in Koopmans DFT}

From a theoretical perspective, the eigenstates of KS-DFT, obtained using a regular XC functional, provide a reasonable first approximation for studying charge excitations in  \cite{hohenberg1964,kohn1965}. Still, while KS-DFT serves us to accurately predict GS properties, its eigenvalues 
$\epsilon_i^{\text{KS}}$ and orbitals 
$\phi_i^\text{KS}(\mathbf{r})$ do not directly describe charged excitations such as the set of vertical ionization energies, i.e., the QPs. This is true even though that all the necessary information to describe a frequency moment of the fully interacting spectral function is, in fact, available 
by simply combing terms of the
KS-DFT energy descriptions with
also an evaluation of the kinetic-correlation energy \cite{Burke97,signatures,ChiDFT23}.
Even a PBE-based evaluation of those DFT terms yields a value for the moment of the interacting spectral function that closely reflects
a full multi-reference configuration integral studies for small molecules
\cite{ChiDFT23}.

The limitation of using KS-DFT orbitals as a predictor for QPs arises because KS-DFT relies on using (regular, nonhybrid) XC functionals 
to define an effective
local potential contribution \cite{helujpc1971,kronik2012} to set the orbital structure. Plotting the total energy as a function of the number of electrons then yields a convex curvature \cite{kronik2012}. Use of regular XC functionals to thus directly set KS-DFT orbitals implies a lack of the PWL that is, for example, needed for a DFT-type framework to also satisfy a thermodynamics constraints on charge transfers \cite{perdew1982,kronik2012}.

Additional information, concealed as MBPT input to the XC functional designs, is needed to also recover accurate results for QP levels \cite{davo2010,kronik2012,WiOhHa21}.
Fortunately the information is still formally retained in the MBPT input and physics logic of good XC designs \cite{ChiDFT23}. By switching to  Koopmans-compliant DFT and using the 
KI-DFT framework with a linear-response extensions of an underlying XC functional (like PBE or the CX version of the vdW-DF method), we can leverage more of the underlying MBPT contents \cite{helujpc1971,hyldgaard2020,ChiDFT23} of such robust XC designs also for accurate QP predictions \cite{colonna2022,schubert2023,OTRSHalga,OTRSHgap,WiOhHa21,JPCM2025}.

On a formal level, KI-DFT employs a class of orbital-density-dependent functionals (ODDFs) designed to enforce \textit{generalized PWL.} 
Specifically, the ODDFs are  
crafted to enforce that we have a linear energy change with respect to variations in the electron occupation \cite{perdew1982,kronik2012,davo2010}.
The CX-based adjustments for KI-DFT, what can also be called the Koopmans functional, is given:
\begin{equation}
    \Ek{KI-CX}{xc}{\rho, \{\rho_{i}\}} = \En{CX}{xc} + \sum_{i} \alpha_{i} \Pi_i^{\text{CX}}[\rho,\rho_i]\, .
    \label{eq:koop}
\end{equation}
Here $\{\alpha_{i}\}$ are screening parameters that account for electronic relaxation effects and $\rho_i$ are 
orbital-specific density variations
$\rho_i(\mathbf{r})=f_i|\phi_i(\mathbf{r})|^2$; with $0<f_i<1$ \cite{davo2010}. 
We are using the KI 
form \cite{borghi2014,nguyen2018,colonna2022,linscott2023} so 
that  the formal KI-functional specification, 
$\Ek{KI-CX}{xc}{\rho, \{\rho_{i}\}}$,  
reduces exactly to the underlying
CX functional form when there is an  integer charge in the unit cell
(and hence KI-CX does not change 
KS-DFT structure predictions obtained by CX).
However, for partial orbital occupation, the Koopmans-correction term $\Pi_{i}^{\text{CX}}$ in Eq.\ (\ref{eq:koop}) restores PWL: It converts the nonlinear dependence of the CX functional on changes in orbital-specific occupations, $\int \rho_i(\mathbf{r})d\mathbf{r}$, into a linear form between any two integer occupations. This approach suffices to convert the KS-DFT input (i.e., the choice of the XC) into the KI-DFT tool for making corresponding QP predictions \cite{nguyen2018,gennaro2022,linscott2023}. Further details are provided in Refs.\ \cite{davo2010,linscott2023}.

Conventionally, spectral properties computed via Eq.~\eqref{eq:koop} require minimization of, e.g., the
$\Ek{KI-CX}{xc}{\rho, \{\rho_{i}\}}$ term,
through a two-step nested procedure. The inner loop minimizes the orbital-density-dependent functional contribution, while the outer loop optimizes orbitals in directions orthogonal to the subspace. At convergence one obtain the variational orbitals. These variational orbitals enable computation of $\qty{\alpha_{i}}$ via finite differences. For periodic systems, this parameterization approach requires supercell treatment to eliminate spurious interactions between periodic images.

Although this method rigorously provides both variational orbitals and $\{\alpha_{i}\}$, its application remains challenging due to the requirement of multiple constrained DFT and Koopmans calculations in supercells for periodic systems. To address this limitation, the \textsc{kcw} formulation, 
discussed in Refs.\  \cite{baroni2001,colonna2022,gennaro2022}, employs a second-order Taylor expansion of $\Pi_i[\rho,\rho_i]$ and use of DFPT in primitive cells. This approach 
assumes coincidence between variational orbitals and maximally localized Wannier functions (MLWFs) $\qty{w_{i\mathbf{k}}}$) \cite{mazari2012}, that are computed by use of the same underlying functional. These MLWFs simplify the  minimization of $\Ek{KI-CX}{xc}{\rho, \{\rho_{i}\}}$.

To obtain spectral properties within \textsc{kcw},  the resulting total energy $\En{KI-CX}{tot}$ is  varied with respect to the orbital densities $\rho_i(\mathbf{r})$. This is done by representing it in the Wannier-function basis set
\begin{equation}
\bra{w_{\mathbf{k}i}} \left( \hat{h}^{\text{CX}} +  \alpha_{\mathbf{0}i}\mathcal{V}^{\text{CX}}_{\mathbf{0}i} \right)  \ket{w_{\mathbf{k}j}} 
=
\mathcal{H}_{ij}\qty(\mathbf{k})
\label{eq:kcw_ham}
\end{equation}
where  $ \mathcal{V}^{\text{KI}}_{i\mathbf{0}}$ is the orbital density dependent potential (ODDP) and corresponds to the variation $\delta\sum_j \Pi_j/\delta \rho_{i}$, and $\hat{h}^{\text{CX}}$ is the effective potential defined by CX functional \cite{burke}. The subscript index $\mathbf{0}$ in $\mathcal{V}^{\mathrm{CX}}_{\mathbf{0}i}$ and $\alpha_{\mathbf{0}i}$ refers to a decomposition into a set of independent problems, where a supercell calculation with $\Gamma$-point sampling corresponds to sampling the folded Brillouin zone of the primitive cell \cite{colonna2022}.
Hamiltonian diagonalization  in Eq.~\eqref{eq:kcw_ham}, 
\begin{equation}
        \mathcal{H}_{ij}(\mathbf{k}) \rightarrow \varepsilon^{}_{\mathbf{k}i} \delta_{ij} \, ,
\label{eq:diag_koopsham}
\end{equation}
yields so-called  canonical orbitals that corresponds to the eigenvectors $\Tilde{w}_{\mathbf{k}i}$. The
eigenvalues of Eq.\ (\ref{eq:diag_koopsham}),
$\varepsilon^{}_{\mathbf{k}i}$ define the KI-CX (or KI-DFT) predictions for the QP energies. We note that the canonical orbitals have a Block-wavefunction-like character; They are delocalized as they are linear combination of MLWFs $w_{\mathbf{k}i}$.
It has been shown that use of an ODDP can provide a  QP approximation to the frequency-dependent contraction of the many-body self-energy term \cite{ferretti2014}. The implication is that ODDFs have the flexibility to accurately describe both total energies and QP excitations.

To elucidate the role of the CX functional in the QP states described by Eq.~\eqref{eq:kcw_ham}, we note the dependence on the XC kernel $f_{\text{Hxc}}$, that is, the double derivative of the XC energy functional with density variations. This kernel includes a contributions from the nonlocal-correlation energy functional defined in Eq.~\eqref{eq:Enl2}, specifically, through its second functional derivative:
\begin{equation}
f_{\text{c}}^{\text{nl}} \qty(\mathbf{r},\mathbf{r}') = \frac{\delta^{2} \En{nl}{c}}{\delta \rho(\mathbf{r}) \delta \rho(\mathbf{r}')} \, .
\label{eq:nl-lrkernel}
\end{equation}
This linear-response kernel $f_{\text{c}}^{\text{nl}}$  depends 
itself on the electronic density $\rho(\mathbf{r})$. This dependence
is in turn explicitly incorporated into the screening parameters $\{\alpha_{i}\}$, thereby influencing the quasi particle energy solutions.

Appendix A provides details on a formal representation of a plane-wave and grid-based determination of
Eq.\ (\ref{eq:nl-lrkernel}) using the Roman-Soler spline formulation
\cite{roso09} for (the functional nature \cite{dion2004,thonhauser2007,hybesc14} of) the nonlocal-correlation kernel $\Phi_c^\text{nl}[\rho](\mathbf{r},\mathbf{r'})$ entering Eq.\ (\ref{eq:nlenergybykernel}).
The formal expressions are already coded in the linear-response module (\textsc{kcw}) of QE thanks to a previous study documenting fingerprints of vdW and nonlocal-correlation interactions  on phonon predictions \cite{sabatini2016}.

\section{Computational details}

We compute the QP states by the \textsc{Koopmans}-DFT code package \cite{koopweb2025,linscott2023}, using the KI-DFT versions,  
and the above-summarized \textsc{kcw} formulation 
\cite{colonna2022}. All those calculations
as well corresponding KS-DFT studies are done using norm-conserving PBE-type (scalar-relativistic) pseudopotentials from the PseudoDojo library \cite{vansetten2018}. That is,
we employ dual projectors and generalized norm conservation to accurately reproduce all-electron potential binding and scattering properties. 
Appendix B presents a discussion of numerical stability for the QP-energy predictions with the wavefunction energy cut off $E_{\rm cut}^{\rm ef}$ (and, for the strictly two-dimensional BN1 system, the height $c$ of the
modeling cell). This analysis leads us to use $E_{\rm cut}^{\rm wf}=200$ Ry (400 Ry) for bulk (single-sheet) h-BN, while doubling the grid for the density variations; Exceptions are explicitly stated. 

The workflow for KI-DFT QP studies can be summarized in  the following steps: 
\begin{enumerate}
    \item Do a MLWFs ($\ket{w_{\mathbf{k}i}}$) for the KS states.
    \item Convert the $\qty{\ket{w_{\mathbf{k}i}}}$ to Wannier orbital densities $\qty{ w_{\mathbf{k}i}^{*}w_{\mathbf{k'}i} }$ 
    \item Calculation of $ \{ \alpha_{\mathbf{k}i}\} $ via DFPT.
    \item Diagonalize the Hamiltonian in Eq.~\eqref{eq:kcw_ham} to obtain $\qty{\Tilde{w}_{\mathbf{k}i}}$ and $\qty{\varepsilon^{}_{\mathbf{k}i}}$
\end{enumerate}

In the initial step, \textsc{Koopmans} code  executes the QE codes \textsc{pw.x} and \textsc{ph.x}. Subsequently, the MLWF algorithm is carried out through \textsc{pw2wannier90.x} and \textsc{wannier90.x}~\cite{pizzi2020}. Steps two through four employ the \textsc{kcw.x} code
that also gave the workflow naming \cite{colonna2022}.

The QP band structure calculations is given by  Eqs.\ \eqref{eq:koop} and \eqref{eq:kcw_ham}; The orbital resulting in KI-CX results are themselves the KI-CX QP predictions. These results are compared to QP states of KI-PBE and the KS states of the corresponding underlying functionals CX and PBE. KI-PBE calculations are standard in the \textsc{Koopmans}-DFT code. To do band structure calculations with KI-CX it is necessary to explicitly introduce a selection of the CX functional in the `QE' specification, see tutorial Ref.\ \cite{koopweb2025}.

\begin{figure}
    \centering
   \includegraphics[width=8cm]{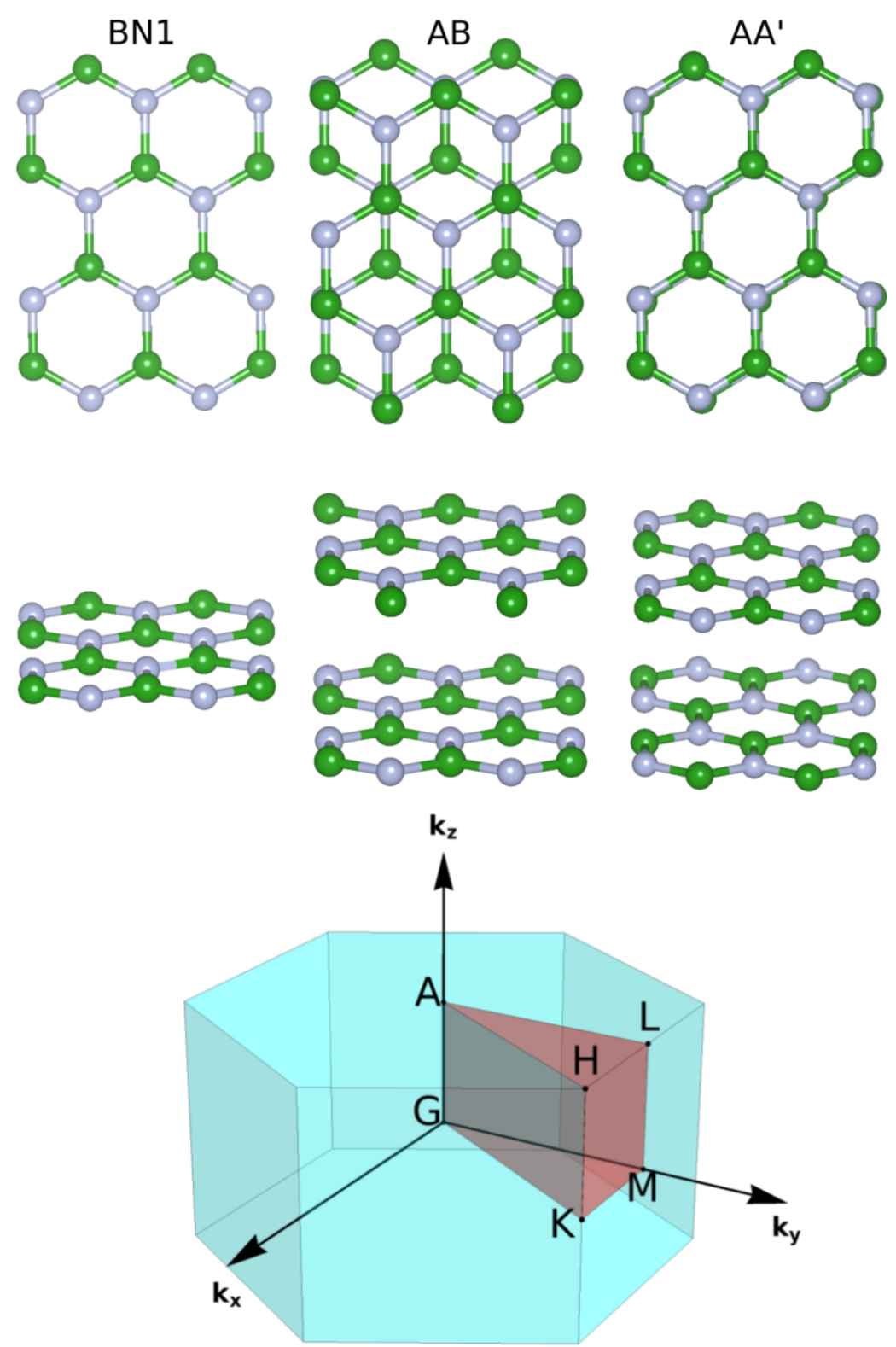} 
    \caption{Atomic as well as Brillouin-zone structure of h-BN structures in the single-sheet form, denoted BN1, and in the two stable bulk forms, AB and AA. The panels of the first and second row 
    show the atomic structures as shown from a perspective perpendicular and parallel to the slide plane. The bottom panel shows the Brillouin zone (BZ) of the bulk h-BN forms in blue and the reduced BZ zone in red. This panel also identifies the set of high-symmetry $k$-points of the reduced BZ as relevant for these hexagonal systems.
     }
    \label{fig:BN_structures}
\end{figure}

We compute QP states KI-CX (technically, `Koopmans DFPT' \cite{colonna2022} with a CX start) for the h-BN monolayer (BN1) and for bulk h-BN in the experimentally observed AB and AA' stacking configurations. The structural characters of these h-BN systems are illustrated in Fig.~\ref{fig:BN_structures}  that
also identifies the set of relevant high-symmetry $k$-points in the Brillouin zone (BZ).

The BN1 structure (first column) is a monolayer exhibiting the conventional \textit{honeycomb} lattice, where boron (green) and nitrogen (gray) atoms with valence 3 occupy alternating vertices of the hexagonal network. In-plane $sp^{2}$ bonding between boron and nitrogen atoms provides structural stability and rigidity for this single-sheet system.  The choice of in-plane lattice constants for the BN1 system is set by an early experimental structure characterizations of h-BN \cite{hassel1927}: For BN1 we use an hexagonal unit cell with in-plane parameters  $a=b=2.504$  {\AA} and a $7\times 7 \times 1$ $k$-point grid
sampling. In our periodic modeling used for QE studies we set $c=20$ {\AA}; This choice gives sufficient distance between repeated images of the BN1 layer that we can, in practice, ignore spurious inter-layer forces in characterizations of cohesion as well as of QP energies, as discussed in Appendix B. 

Bulk AB stacking (second column of
Fig.\ \ref{fig:BN_structures}) features a graphite-like staggered 
arrangement of hexagons where half of the atoms (for example, all
of the nitrogens) sit directly on top of a different type of 
atom in neighboring layers.
For our AB calculations, we use a  $7\times7\times3$ $k$-point grid with lattice constants set at an average over measurements on what are interpret \cite{lifesh2003} as experimentally 
characterized AB structures, $a=b=2.505$ {\AA} and $c=6.661$ {\AA}.

The third column of panels in Fig.\ \ref{fig:BN_structures} show 
schematics of the AA$'$ form that is the ground-state of h-BN. 
Contrary to the AB form, the ground-state configuration stacks the 
BN layers so that all of boron (nitrogen) atoms in a given layer
sit directly above a nitrogen (boron) atom in neighboring layers. 
For our AA' studies, we use an hexagonal unit cell with lattice constants, 
$a=b=2.504$ {\AA} and $c=6.66$ {\AA}, that represent averages of 
values extracted by experiments
\cite{hassel1927,pease1952,lifesh2003}. 
Also, we again use a  $7\times7\times3$ $k$-point sampling. 

Additionally, we include a study demonstrating
the full vdW impact on the QP band structure, contrasting what we call consistent-KI-CX and consistent-KI-PBE descriptions. That is, we track not only the \textit{direct\/} changes arising when we switch from a PBE to a CX starting point in KI-DFT at the same fixed geometry, but also the \textit{additional indirect signatures} that arise because a GGA PBE is not set up to accurately characterize the bulk h-BN structure \cite{rydberg03p606,rydberg03p606,rohrer11p165423,hyldgaard2020}. The importance of this is discussed in the following section IV. In practice, we complete 
KI-PBE and KI-CX QP characterizations, having first relaxed the interlayer distance of the AA$'$ system using KS-DFT with the PBE and the CX functionals, respectively. Specifically, we find that 
CX-based and PBE-based structure optimizations yield lattice constants $c=6.448$ {\AA} and $c=8.560$ {\AA}, 
respectively. The former and latter structure is then used
in ensuing KI-CX(CX) and KI-PBE(PBE) QP characterizations. 

For the three materials forms at experimental structures, and for the 
demonstration of indirect vdW fingerprints,
we presents analysis of the resulting QP bandstructures, using a shared computational treatment.  The Gygi-Baldereschi method~\cite{gygi1986} is used to address the $\mathbf{q}\to 0$ divergence of the Coulomb interaction in our periodic systems, both in the initial QE code stages and when setting the screening coefficients $\alpha_i$ for the 
KI-DFT studies, above. The Gygi-Balderashi
approach assumes some level of isotropy and 
its use is more motivated for the bulk BN cases than for the single-sheet BN1 system.

As highlighted in Eq.~(\ref{eq:kcw_ham}), Wannier functions serve as the initial basis for our set of \textsc{Koopmans} DFPT studies. That is, we use MLWFs  to generate localized orbitals $\qty{\ket{w_{\mathbf{R}i}}}$ and we find a set of initial Wannier functions that work for the layered h-BN systems. These initial Wannier functions are selected using an initial projected density of states (PDOS) that we obtain via use of the QE \textsc{ projwfc.x} code. They take the form of modified spherical harmonics, with details  listed in Table \ref{tab:proj}. 

The occupied bands are located at the nitrogen atoms in the unit cell, with the following angular momentum shapes $s$  and $p_x$, $p_y$, and $p_z$ and each of these having a radial function $R_2\qty(r)$. For the virtual states we chose $p_z$ but now located at the boron atoms, with the radial functions $R_2\qty(r)$. This process is verified by the convergence of the MLWF algorithm. We have also tried to use hybrid $sp2$ orbitals instead but they did not converge in the MLWF procedure, even though suggested by chemical intuition.

\begin{table}
    \caption{Projections for the iterated MLWF determination.}
    \centering
\begin{tabular}{c c  c } % @{} removes padding
    \hline
     Site & Angular momentum $\Theta\qty(\theta,\phi)$ & Radial functions    \\
    \hline
    \multicolumn{3}{c}{Occupied bands}\\
    \hline
    N     & s & $R_2\qty(r)$ \\
    N     & $p_x$, $p_y$, $p_z$& $R_2\qty(r)$ \\
    \hline
    \multicolumn{3}{c}{Virtual bands}\\
    \hline
    B     & $p_z$ & $R_2\qty(r)$ \\
     \hline
\end{tabular}

    \label{tab:proj}
\end{table}

Beyond initial Wannier-function selection, we must disentangle virtual bands from occupied states to construct well-localized Wannier functions. The disentanglement window spans from the Kohn-Sham highest occupied molecular orbital (HOMO) to \texttt{dis\_win\_max}, which defines the upper energy bound for disentanglement. Simultaneously, \texttt{dis\_froz\_max} sets the upper limit of the frozen energy window. 
The layered bulk systems AA$'$ and AB we employ parameter values of \texttt{dis\_froz\_max} = 0.6\,eV and \texttt{dis\_win\_max} = 16.0\,eV. For the monolayer BN1 system, we adjust these parameters to \texttt{dis\_froz\_max} = 0.5\,eV and \texttt{dis\_win\_max} = 14.0\,eV.

In contrasting predictions of the KI-DFT predictions and KS-DFT approximations
for the QP bandstructures we face, in principle, a problem of how to pick a reference energy. Bandstructure characterizations are 
often reported with QP energies listed relative to the valence band maximum (VBM) but such values are generally affected by both the atom configurations and choice of methodology. Ideally, as the QPs 
reflect vertical excitations, we would
like to list the variation with QP energies defined relative to the true vacuum. However, in a planewave code like QE, we can only get orbital-energy values relative to what is the average electrostatic potential in a given cell, with a given atomic configuration (and choice of pseudopotentials) and for the electron-density variation that is produced
in the given DFT method and given choice of XC functional \cite{RMBook}. 

\begin{figure}
    \centering
    \includegraphics[width=1\linewidth]{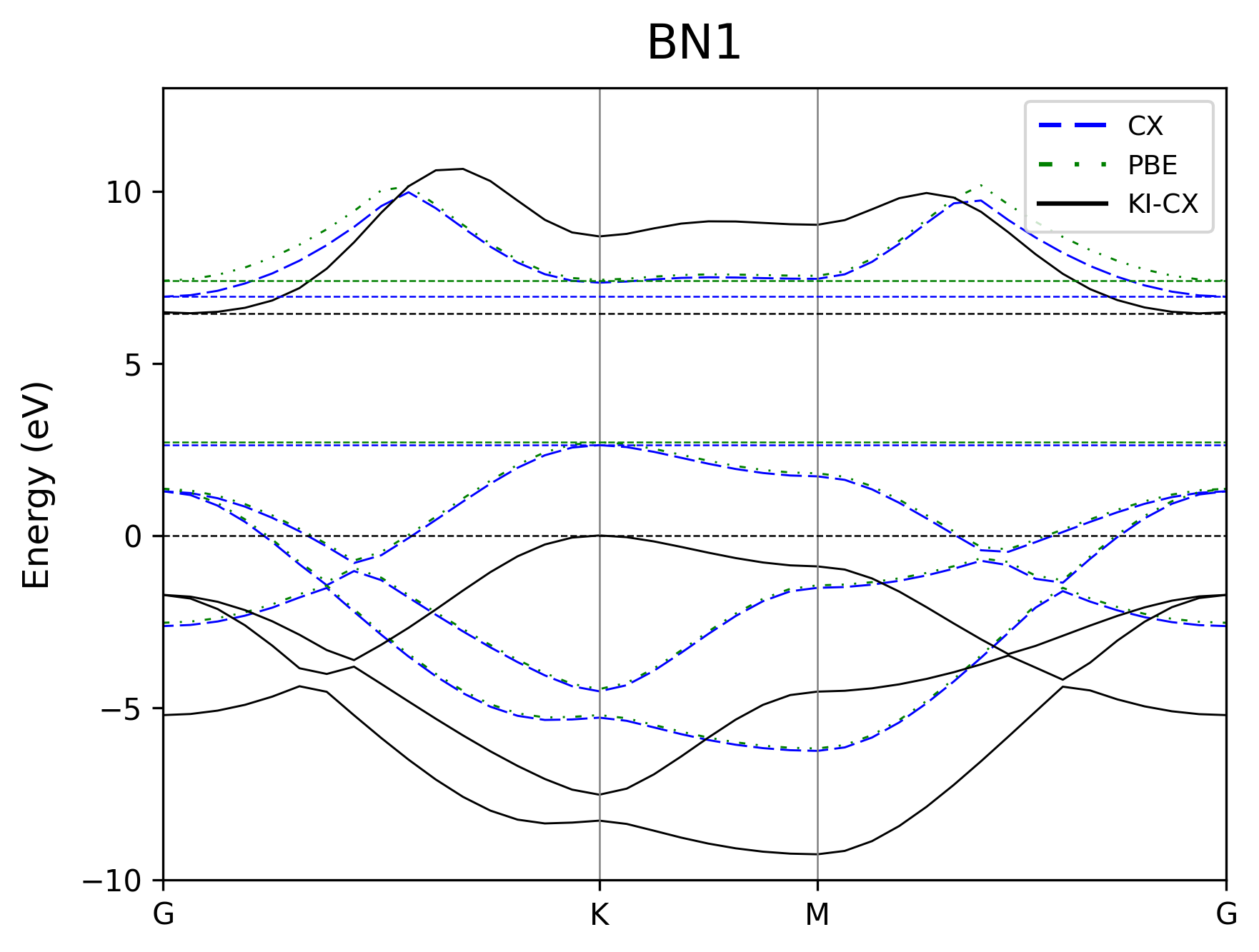}
    \caption{Comparison of BN1 QP band-structure predictions, as evaluated in PBE, CX and in KI-CX in a $c=20$ {\AA} cell to minimize the impact of spurious couplings between repeated images in our planewave-DFT studies. 
    The top of the KI-CX QP-valence band is used as a shared reference energy, as motivated in the text. 
}
    \label{fig:bn1_bands}
\end{figure}

Nevertheless, in predictions and discussions of the QP bandstructure differences \textit{at fixed experimental structures,} it is possible and motivated to define a common reference. Specifically, in 
Figs.\ \ref{fig:bn1_bands} and \ref{fig:main_bands}  we have shifted all the bands using
\begin{equation}
 \sum_{\mathbf{k},i} \varepsilon'^{M}_{\mathbf{k}i} = \sum_{\mathbf{k},i}\varepsilon^{M}_{\mathbf{k}i}-  \varepsilon^{KI-CX}_{\texttt{VBM}} 
\end{equation}
where  $\varepsilon^{KI-CX}_{\texttt{VBM}} $ the VBM of KI-CX and where $M$ corresponds to difference choices of the methodologies: PBE, CX, KI-PBE and KI-CX. The point is that since the atomic and cell geometry are shared (and we use the same pseudopotentials in all studies) the average electrostatic potential can only vary because of differences in the electron-density variation. However, PBE and CX are both widely tested and both documented to provide highly accurate accounts of the electron-density variations (at fixed structure) that must therefore be almost identical in the case of PBE and CX studies. Furthermore, by design of the KI-DFT
method, it is also so that the KI-CX and CX yields exactly the same electron density variations (at shared geometries) and the same applies for KI-PBE and PBE \cite{linscott2023}. By the
explicit procedure that the QE code uses to set the average electrostatic potentials, Refs.\ \cite{Tosi1964,RMBook}, no differences
can therefore arise for the electrostatic reference energy in KI-CX and CX. 
The \textit{same offset} brings the here-reported bandstructures in alignment with a QP bandstructure mapping that reflects the true vacuum level.

\section{Results and Discussion}

Unlike graphite, h-BN experimentally demonstrates five distinct stacking arrangements, with only two being thermodynamically stable and the remaining three metastable \cite{quian2024,ooi2006}. A key factor driving interest in h-BN is its wide band gap, which is systematically underestimated in KS-DFT \cite{ooi2006}. In contrast, GW approximations align more closely with experimental observations \cite{blase1995}. Here we seek to compute and contrast QP band structure results of the stable h-BN forms AA'-stacked, AB-stacked, and monolayer h-BN (BN1),  Fig.~\ref{fig:BN_structures}.

Our study and analysis highlight changes arising as we move from KS-DFT to KI-DFT and as we include (in KI-CX) nonlocal-correlation effects.
In particular, we track differences (i) between use of semi-local and truly nonlocal-correlations in KS-DFT (PBE vs.\ CX results); (ii) between use of a semi-local functional for KS-DFT and for its KI-DFT counterpart (PBE vs.\ KI-PBE); (iii) between use of a vdW-DF for KS-DFT and for the corresponding KI-DFT (CX vs.\ KI-CX); and finally, (iv) arising when we include truly nonlocal-correlations in KI-DFT (comparing KI-CX and KI-PBE results).

\subsection{Bandstructure results and analysis}

\begin{figure*}
    \centering
        \includegraphics[width=17.5cm]{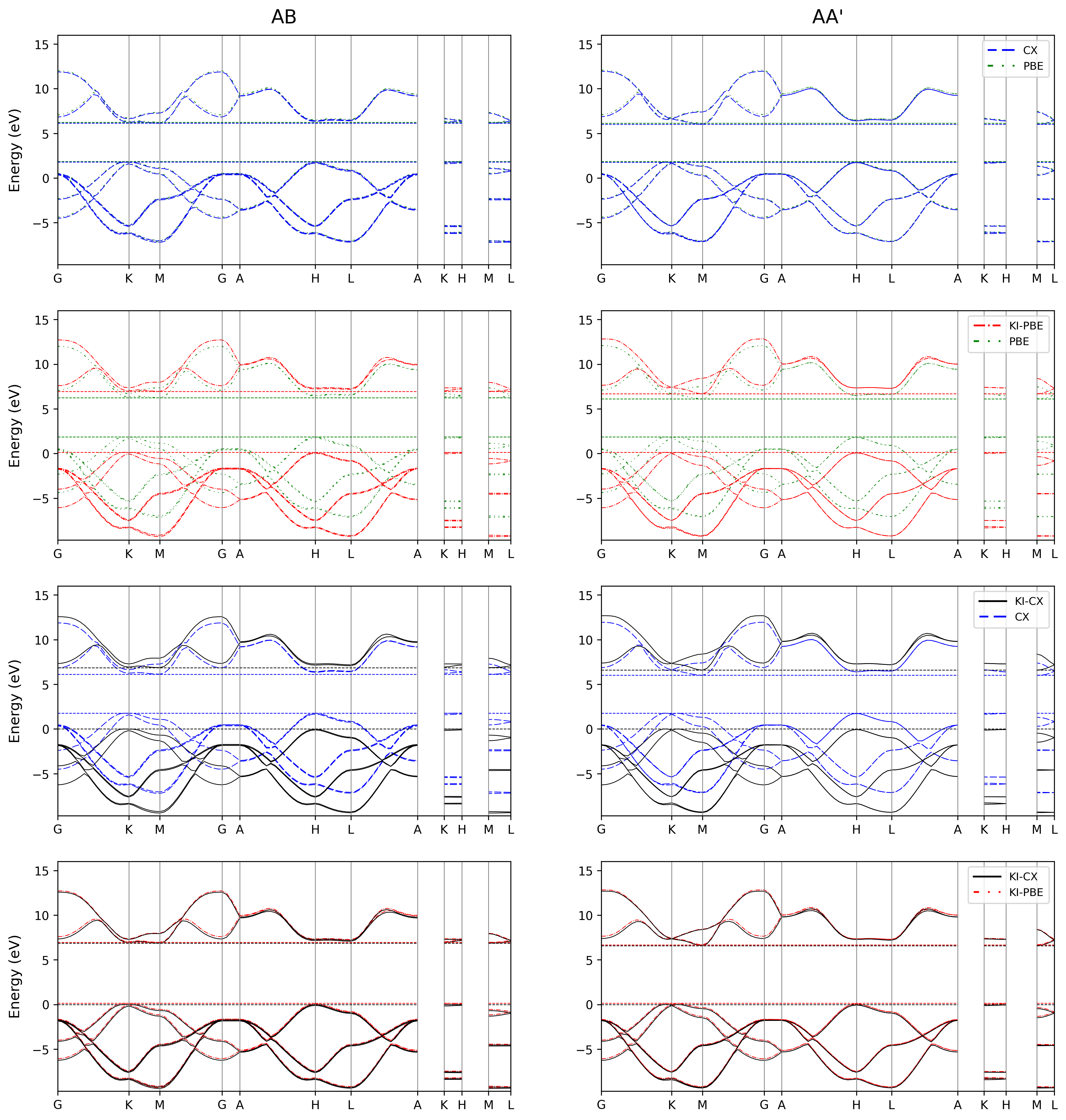}
        \caption{Band structure comparison for h-BN structures AB and AA$'$, with KI-PBE and KI-CX results provided in KI-DFT and PBE and CX results in KS-DFT, i.e., using the same underlying CX functionals.
     }
    \label{fig:main_bands}
\end{figure*}

\begin{table*}
    \centering
    \begin{tabular}{l c c c c c c c}
    \hline
    Method & $\text{G}_{v}\rightarrow \text{G}_{c}$ & $\text{M}_{v}\rightarrow \text{M}_{c}$ & $\text{K}_{v}\rightarrow \text{G}_{c}$ & $\text{K}_{v}\rightarrow \text{K}_{c}$ & $\text{M}_{v}\rightarrow \text{G}_{c}$ & $\text{K}_{v}\rightarrow \text{M}_{c}$ & Fundamental Gap \\
    \hline
    PBE & 6.04 & 5.74 & 4.69 & 4.71 & 5.59 & 4.83 & 4.69$^{a}$ \\
    CX & 5.65 & 5.74 & 4.31 & 4.72 & 5.22 & 4.83 & 4.31 \\
    KI-CX$^b$   & 8.21 & 9.92 & 6.48 & 8.69 & 7.38 & 9.02 & 6.48 \\
    KI-CX$^c$  & 8.20 & 9.99 & 6.47 & 8.72 & 7.37 & 9.09 & 6.47 \\
    $GW_0^{d}$  & - & - & 7.43 & 7.90 & - & 8.00 & 7.43 \\
    \hline
    \multicolumn{8}{l}{$^{a}$ Nearly degenerate.} 
    \\
    \multicolumn{8}{l}{$^{b}$ Computed at $c=20$ {\AA}, at $E_{\rm cut}^{\rm wf}=400$ [Ry].}
    \\
    \multicolumn{8}{l}{$^{c}$ Computed at $c=30$
    {\AA}, at $E_{\rm cut}^{\rm wf}=400$ [Ry].}
    \\
        \multicolumn{8}{l}{$^{d}$ Ref.\ \onlinecite{hunt2020}.} 
    \\

        \hline
    \end{tabular}
    \caption{QP-gap energies (in eV) by PBE, CX, and KI-CX for valence-to-conduction-band transitions at different symmetry points for the single-sheet h-BN structure BN1, with comparison to literature $GW_0$ results. 
    }
    \label{tab:bn1}
\end{table*}

In Figs.\ \ref{fig:bn1_bands} and \ref{fig:main_bands}  we show comparisons of the band structure predictions obtained at experimental structures by our set of methodologies.

Figure \ref{fig:bn1_bands} summarizes our results for the single-sheet BN1 case
(at an experimentally relevant atomic 
in-plane geometry). Here we trace out the QP dispersion along 
a $k$-point path defined by the string 
`GKMG' of high-symmetry points that are relevant for a two-dimensional hexagonal system, see bottom panel of Fig.\ \ref{fig:BN_structures}.
 
Figure \ref{fig:main_bands} presents our 
bandstructure predictions obtained for the two h-BN bulk systems (AB and AA').
There we trace out the QP-energy variation along a richer path, `GKMGAHLA', of the high-symmetry $k$-points that are relevant for such bulk hexagonal systems, see again the bottom panel of Fig.\ \ref{fig:BN_structures}.
The panels in the first (second) column
of Fig.\ \ref{fig:main_bands}, report our results for bulk h-BN in the AB (AA') form. The first row of panels presents
comparisons of the QP 
bandstructure approximations defined by KS orbitals
and obtained in PBE and CX, for the two bulk structures. The third (second) row of panels contrasts QP predictions as provided by CX and KI-CX (by PBE and KI-PBE) studies. The last row of panels reports comparisons of KI-PBE and KI-CX QP predictions, again for the two bulk h-BN forms.

We note in passing, that we are not showing the internal (2s) first bands
in any of our band structure mappings.
This is done to present a clearer view of what is happening near the valence and conduction band edges. 

In the following discussion of the band structures, we will use subscripts at symmetry points to denote the valence- and conduction-bands that define various (generally indirect) gaps of potential key interband transitions. For example, $\text{K}_{c}$ indicates the conduction band at the K point, while $\text{K}_{v}$ denotes the valence band at the same symmetry point.  The set of key (potential) interband transitions are identified by the start and end points,  e.g., $\text{K}_{v}\rightarrow \text{K}_{c}$, when we summarize and discuss our set of QP predictions across the structures.

\subsubsection{BN1}

In Fig.~\ref{fig:bn1_bands} we compare PBE, CX and KI-CX results for the QP band structure in the monolayer BN1 system, here in the standard
set up using a $c=20$ {\AA} cell height for the
computational modeling.  Table~\ref{tab:bn1} provides a compilation of our quantitative results for gaps defined
by valence-to-conduction-band transitions among the set of high-symmetry $k$-points. 
We note that we could not complete a corresponding KI-PBE study because the KI-DFT computational scheme fails to converge  the screening parameters that enter in the KI-PBE functional for conduction-band QPs.

Contrasting the KS-DFT descriptions by PBE and CX, we observe a small (a fraction of an eV) rigid downward shift for the valence bands in CX relative to PBE. In the conduction band, CX shifts states downwards occurring at $\text{G}_{c}$, resulting in a slightly reduced band gap. In both cases, the VBM resides at the $\text{K}_{v}$ symmetry point. The crucial difference lies in the conduction band minima (CBM): PBE finds the CBM as
essentially degenerate between $\text{K}_{c}$ and $\text{G}_{c}$, predicting a minimum direct quasiparticle band gap (QP-gap) $\text{K}_{v}\rightarrow \text{K}_{c}$ of 4.71 eV 
and minimum indirect gap $\text{K}_{v}\rightarrow \text{G}_{c}$ of 4.69 eV.
In contrast,  CX clearly indicates the indirect transition $\text{K}_{v}\rightarrow \text{G}_{c}$ as the fundamental gap at 4.31 eV (compared to 4.72 eV for $\text{K}_v \to \text{K}_c$), in qualitatively agreement with $GW_0$ results, see  Table~\ref{tab:bn1}.

Comparison of KI-CX with CX results reveals an essentially rigid downward shift of about 2.5 eV in the KI-CX valence bands. For the conduction band we see that KI-CX lowers the $\text{G}_{c}$ QP energy, while $\text{M}_{c}$ and $\text{K}_c$ QP energies are shifted upwards. As a consequence, we find that use of KI-CX as a QP predictor does provide an adjustments of the underestimated PBE and CX gaps.

The left panel of Fig.\ \ref{fig:conduct_band_minimum}
shows a plot of the QP wavefunction as obtained 
by KI-CX for the
conduction-band at the 
high-symmetry $\Gamma$ point `G'. 
This variation is shown for reference. It allows us to
illustrate and discuss some of the impacts that the interlayer vdW attraction has on the details of bulk-AB and bulk-AA$'$ QPs, next.

\begin{table*}
    \centering
    \begin{tabular}{l c c c c c c c}
    \hline
    Method & $\text{G}_{v}\rightarrow \text{G}_{c}$ & $\text{M}_{v}\rightarrow \text{M}_{c}$ & $\text{K}_{v}\rightarrow \text{G}_{c}$ & $\text{K}_{v}\rightarrow \text{K}_{c}$ & $\text{M}_{v}\rightarrow \text{G}_{c}$ & $\text{K}_{v}\rightarrow \text{M}_{c}$ & Fundamental Gap \\
    \hline
    \multicolumn{8}{c}{AB} \\
    \hline
    PBE & 6.53 & 5.10 & 5.22 & 4.51 & 5.91 & 4.41 & 4.39$^{a}$ \\
    CX & 6.36 & 5.07 & 5.07 & 4.49 & 5.76 & 4.38  & 4.36$^{a}$ \\
    KI-PBE & 9.23 & 7.52 & 7.48 & 6.92 & 8.14 & 6.86 & 6.83$^{b}$\\
    KI-CX  & 9.06 & 7.56 & 7.36 & 6.96 & 8.01 & 6.90 & 6.87$^{b}$\\
    HSE+D3$^c$ & - & - & - & 6.23 & - & - & 6.02$^{b}$ \\
    \hline
    \multicolumn{8}{c}{AA'} \\
    \hline
    PBE & 6.58 & 4.75 & 5.35 & 4.95 & 5.71 & 4.39 & 4.28$^{b}$\\
    CX & 6.41 & 4.71 & 5.20 & 4.96 & 5.55 & 4.36 & 4.24$^{b}$\\
    KI-PBE & 9.30 & 7.02 & 7.62 & 7.39 & 7.97 & 6.67 & 6.54$^{b}$\\
    KI-CX & 9.14 & 7.06 & 7.50 & 7.46 & 7.85 & 6.71  & 6.58$^{b}$ \\
    KI-CX(CX) & 8.65 & 6.88 & 7.06 & 7.49 & 7.30 & 6.64 & 6.47$^{b}$ \\
    $GW^c$ & 8.40 & 6.50 & 6.90 & 6.90 & 7.30 & - & 5.95$^{b}$\\
    Exper.$^d$   & - & - & - & - & - & - & \textbf{6.08} \\
    \hline
    \multicolumn{8}{l}{$^{a}$ Corresponds to  $\text{H}_{v}\rightarrow \text{M}_{c}$} \\
    \multicolumn{8}{l}{$^{b}$ Corresponds to  $\text{T}_{v}\rightarrow \text{M}_{c}$} \\     
    \multicolumn{8}{l}{$^{c}$ Ref.\ \onlinecite{hunt2020}.} \\
        \multicolumn{8}{l}{$^{d}$ Ref.\ \onlinecite{arnaud2006}.} \\
        \hline
    \end{tabular}
    \caption{QP gap predictions (in eV) for valence-to-conduction-band transitions in h-BN bulk structures AB, and AA$'$. Our results computed in the Koopmans-corrected functionals KI-PBE and KI-CX, as well as in the underlying KS-DFT functionals PBE and CX, are compared to literature $GW$ values and experiment, where available.}
    \label{tab:quasiparticle_gaps}
\end{table*}

\subsubsection{AB}

In the first column of Fig.~\ref{fig:main_bands} 
(identified by an `AB' header), we show comparisons of the QP dispersion relations for bulk h-BN in the AB structure. The top section of Table \ref{tab:quasiparticle_gaps}
shows a numerical comparison of
the gap predictions that we extract for the set of valence-to-conduction QP band gaps.

In the first row of the AB column of Fig.~\ref{fig:main_bands}, contrasting PBE and CX results, we observe that the valence bands are nearly identical by either of these KS-DFT descriptions. There is merely a small shift of $\approx$ 0.01 eV downward in CX  near $\text{G}_{c}$. The VBM is  located at $\text{H}_{v}$.  The indirect QP-gap reflects a transition  $\text{H}_{v}\rightarrow \text{M}_{c}$ at 4.39 eV and 4.36 eV for PBE and CX,  correspondingly. 

The second-row panel of AB results shows the comparison of PBE and KI-PBE. The occupied bands of KI-PBE are down-shifted relative to PBE, while the unoccupied bands are shifted higher in energy in KI-PBE. In KI-PBE, the fundamental band gap is indirect and corresponds to the transition $\text{T}_{v}\rightarrow \text{M}_{c}$ at 6.83 eV. Here $\text{T}_{v}$ is a non-symmetric $k$ point that lies on the path from $\text{G}_{v}$ to $\text{K}_{v}$, being close to $\text{K}_{v}$. This AB-system VBM corresponds (as a shifted $\text{K}_{v}$ point) to that which was previously predicted for the AA$'$ system by use of HSE+D3 \cite{wickramaratne2018,HSE06,grimmeD3}. We also observe that degeneracy of the unoccupied bands breaks between $\text{A}_{c}$ - $\text{H}_{c}$ and $\text{L}_{c}$ - $\text{A}_{c}$ in the KI-PBE description.

In the third-row panel, comparing KI-CX and CX results, we find that use of KI-CX again  produces an overall downshifting of the valence-type QP bands while in the conduction bands we observe some differences. Like for KI-PBE, 
in the upper part of the Brillouin-zone hexagon (spanned by high-symmetry points A-H-L-A, figure \ref{fig:BN_structures}), we  observe a breaking of the degeneracy between $\text{A}_{c}$ - $\text{H}_{c}$ and $\text{L}_{c}$ - $\text{A}_{c}$. The results of the second- and third-row panels, taken together, suggest that the use of KI-DFT 
and the enforcement of PWL (rather than the choice of underlying functional) is responsible for breaking the degeneracy. KI-CX predicts the $\text{T}_{v}\rightarrow \text{M}_{c}$ transition as the fundamental gap, at 6.87 eV.

 \begin{figure*}
    \centering
    \includegraphics[width=15cm]{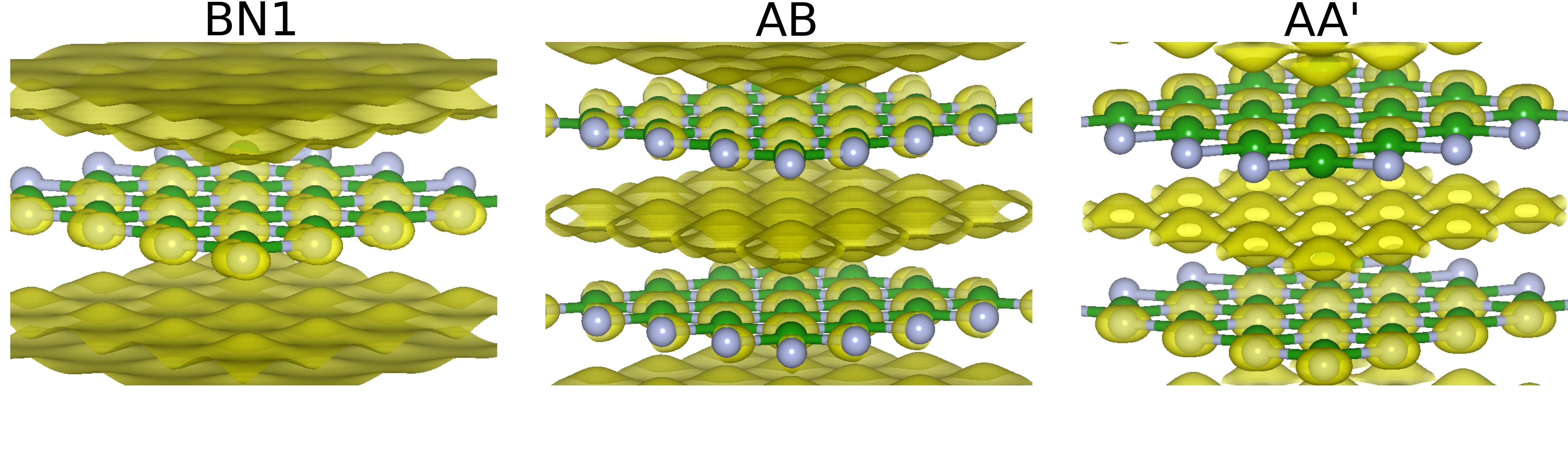}
        \caption{ Real-space canonical orbital densities $\abs{\langle \mathbf{r}  \ket{\widetilde{w}_{\mathbf{R}i}}}^{2}$ at the CBM for h-BN structures BN1, AB, and AA$'$, computed using Koopmans-DFT as KI-CX. The iso-surface constant for BN1 is 5 meV and for AB and AA$'$ corresponds to 7 meV.}
    \label{fig:conduct_band_minimum}
\end{figure*}

In the last-row panel of AB results, we compare the Koopmans methods directly and find that use of KI-CX gives predictions for valence and conduction bands that primarily causes a small downward shift, largest around $\text{G}_{c}$. For the conduction bands, between $\text{A}_{c}$ - $\text{H}_{c}$ and $\text{L}_{c}$ - $\text{A}_{c}$, we also observe a slightly smaller separation between the bands in KI-CX near $\text{A}_{c}$.  

The second panel of Fig.~\ref{fig:conduct_band_minimum} shows the canonical orbital density predicted by KI-CX at the CBM. Contrasting with the corresponding BN1 panel, we find that there is now a different $p_z$-orbital character reflecting orbital overlaps midway between the layers. For the
small $\sigma$-orbital signature of the VBMs,  centered on the nitrogen atoms, there are
also differences.  These VBM changes arise because  the vdW forces provide interlayer attraction to shorten the interlayer separation from essentially decoupled to 
the experimental value that characterizes our AB study \cite{rydberg03p606,rydberg2003,Berland_2015:van_waals,shukla2022}. As such, the QP CBM changes  represent vdW signatures but they should be seen as indirect \cite{rohrer11p165423}, set exclusively by the structural change. In fact, use of the KI-CX and KI-PBE 
(not shown in Fig.~\ref{fig:conduct_band_minimum}) at the experimental structure yields almost identical details in QP CBM orbital densities.

\begin{figure*}
    \centering
    \includegraphics[width=17.5cm]{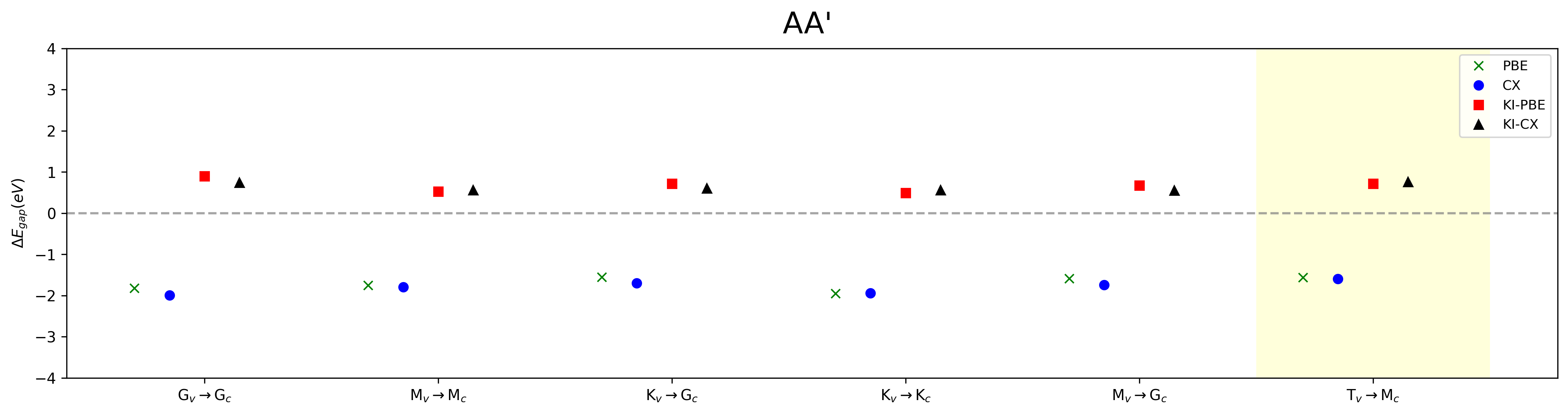}
        \caption{Deviations with respect to $GW$ of our set of predictions for QP-transitions between the indicated set of high-symmetry $k$-points for AA$'$. We highlight the  performance comparison at the fundamental gap,  that is, for the QP transition that corresponds to having the smallest valence-conduction-band QP energy separation.
     }
    \label{fig:error_all}
\end{figure*}

\subsubsection{AA'}
The last column of panels of Fig.~\ref{fig:main_bands} reports our predictions for the QP bandstructure of the AA$'$ system, as described in our set of functionals and DFT frameworks. The bottom section of Table \ref{tab:quasiparticle_gaps}
provides a comparison of our numerical results for the set of high-symmetry band gaps
among the methods and with literature GW characterizations and experiment.

In the first-row panel of AA$'$ results in Fig.~\ref{fig:main_bands}, we find a  similar behavior as in the AB system. The PBE and CX results show no significant differences beyond a small displacement of valence bands, but some changes emerge for the virtual bands. The fundamental gaps emerging in PBE and CX studies reflect again an indirect $\text{T}_{v}\rightarrow \text{M}_{c}$ transition \cite{wickramaratne2018}, at 4.28 eV and 4.24 eV,
respectively. We find that the VBM is at the same non-symmetric $k$ point $\text{T}_{v}$ for the AA$'$ system as for the AB system.

In the second-row panel of AA$'$ results, contrasting PBE and KI-PBE, we observe characteristics similar to those arising for the AB structure. 
The main difference from the AB case is that here KI-PBE reduces the gap by lowering the energy at the symmetry point $\text{M}_{c}$.
 KI-PBE predicts the fundamental-gap QP transition as $\text{T}_{v}\rightarrow \text{M}_{c}$, at 6.54 eV.

In the third-row panel, we contrast CX and KI-CX predictions for the QP band structure. As a 
Koopmans functional, KI-CX predicts what mostly amounts to an overall lowering of the occupied (valence-type) QP energies, relative to the CX description. However, we find that along the path from 
$\text{M}_{v}$ to $\text{G}_{v}$, the band crossing occurs
closer to $\text{G}_{v}$ in KI-CX than for CX.
A similar shift of band crossing is also evident between $\text{L}_v$ and $\text{A}_v$.

Meanwhile, for the conduction-type QP states (unoccupied orbitals), we find that along
the full path from $\text{G}_{c}$ to $\text{A}_{c}$, the use of KI-CX produces a general shift (as does KI-PBE), resembling what is documented 
above for the AB system.
Nevertheless, the switch from 
CX to KI-CX also causes an 
enhanced lowering of the 
conduction-type QP energies
close to $\text{M}_{c}$.
The KI-CX predicts (like KI-PBE) the transition $\text{T}_{v}\rightarrow \text{M}_{c}$ \cite{wickramaratne2018} as the fundamental gap, now at 6.58 eV.

Finally, we compare the two KI-DFT descriptions of the QP bandstructure in the last-row panel of AA$'$ results. We find for the occupied orbitals that use of KI-CX slightly shifts the bands downward while otherwise maintaining a close similarity to KI-PBE results. Both functionals locate the VBM at $\text{T}_{v}$. Again, both methods predict that the fundamental gap corresponds to a QP transition $\text{T}_{v}\rightarrow \text{M}_{c}$,
Table \ref{tab:quasiparticle_gaps}.

Nevertheless that are some differences. In the upper Brillouin-zone region (tracing the set of `A-H-L-A' high-symmetry $k$ points of the BZ), we observe  that KI-CX predicts a small but systematic and visible lowering of QP energies compared with KI-PBE. The overall lowering is expected since the reference value of the potential (by which we measure QP energy positions) is affected by the actual electron-density variation $\rho(\mathbf{r})$. While the CX and PBE densities are indeed so close that CX and PBE provide nearly identical workfunction predictions for transition metals \cite{hyldgaard2020}, there are small 
differences in the $\rho(\mathbf{r})$ profiles
that we compute in CX (and hence in KI-CX) and in 
PBE (and hence in KI-PBE).

The third panel of Fig.~\ref{fig:conduct_band_minimum} identifies again signatures
of the vdW interlayer attraction on the variation of the QP-CBM density evaluated at the $\text{M}_{c}$ point.  As for the AB structure, the KI-CX predictions for the CMB is a QP orbital which exhibits lobes with $p_z$ and $\pi$-bond characters that resemble those found for the
AB structure. However, our KI-CX QP characterization for AA$'$ shows that the corresponding CBM orbital density exhibits a modulation, it has periodically voids in the regions between the h-BN layers. These voids are found at positions that correspond to the centers of the h-BN-atom rings. 
In effect, we find that the CBM QP is trapped along bonds between the hexagonal network of boron and nitrogen atoms.
Accordingly, there are also 
differences at the in-layer $\sigma$-bond positions.

Again we interpret these vdW-binding signatures as indirect 
QP effects because we find essentially no difference in behavior in the KI-PBE characterization of the AA$'$ system (not shown).
We repeat that the PBE lacks an account of vdW and other truly nonlocal correlation effects \cite{hyldgaard2020} and
that KI-PBE cannot provide a full characterization of the QP signatures in 
bulk h-BN. The alignment of KI-CX and KI-PBE characterizations of QPs discussed so far only results because we consider bulk systems with the layer separations fixed at experimentally observed values. We shall return to more detailed arguments for seeing changes in the QP CBM spatial variation relative to the BN1 case, Fig.\ \ref{fig:conduct_band_minimum},
as indirect vdW signatures.

\subsection{Comparison with MBPT}

\textit{Accuracy for BN1 QPs.} Table \ref{tab:bn1} presents an overview of the quantitative results we find in the KI-CX characterization for BN1. We observe that both PBE and CX give gaps that are consistently smaller than those found in $GW_0$ \cite{hunt2020}. While minimal differences exist between these functionals, PBE does yield a better agreement for the BN1 fundamental gap, at the $\text{K}_{v}\rightarrow \text{G}_{c}$ transition, while CX is in closer qualitative agreement with $GW_0$ in that it lifts the PBE 
near-degeneracy with the gap for the $\text{K}_v \rightarrow \text{K}_c$ transition. The descriptions of the QP energies can be seen as clearly improved in KI-CX.

Appendix B summarizes some challenges that we face in numerically converging these KI-CX results with respect to modeling parameters for BN1, for example, forcing us to use a higher $E_{\rm cut}^{\rm wf}$ for the conduction-band QP energies than what is required to converge the KS-DFT/KI-DFT  total-energy description (or valence-band QP energies). We expect this numerical challenge arises, in part, because the screening assumption \cite{gygi1986} that we presently need to use is not naturally set up to handle strictly two-dimensional problems. That is, future KI-CX (or KI-PBE) studies, using a better screening handling for two-dimensional systems \cite{Sohier2017}, may permit computationally cheaper QP descriptions.

\textit{Accuracy for AA$'$ QPs.}  
Table \ref{tab:quasiparticle_gaps} presents
an overview of the accuracy that exists in PBE, CX, KI-PBE, and KI-CX predictions for the QP
energies of bulk h-BN, our main focus. 
The assessments are made in terms of band gaps, that is, energy differences between QP transitions at the set of $k$ points that correspond to local extrema in the QP band structure. Specifically, we make quantitative comparisons of QP predictions for these band gaps by comparing with literature $GW$ results that exist for the bulk AA$'$ structure \cite{hunt2020}. We note in passing that the $GW$ value for the AA$'$-system fundamental gap is in close agreement with an experimental observation \cite{arnaud2006}, see Table  \ref{tab:quasiparticle_gaps}.
We also note that our screening approximation \cite{gygi1986} is motivated for bulk systems and that
we found these calculations robust already when picking the same $E_{\rm cut}^{\rm w}=200$ Ry value that suffices for the total-energy description, as discussed in Appendix B.

Figure \ref{fig:error_all} 
presents an overview of
method performance, as revealed by accuracy on the description compared with $GW$ results that we take as references. The figure focus on the gaps of the AA$'$ system and we mirror the result organization used in Fig.~\ref{fig:BN_structures}. That is, we use identical color coding for the respective KS-DFT and KI-DFT results, summarizing the method performance in terms of the deviation:
\begin{eqnarray}
\Delta E_{\rm gap}^{\alpha,\beta} & = & 
\delta\varepsilon^{\alpha}
-\delta\varepsilon^{\beta} \, ,
\\
\delta\varepsilon^{\alpha/\beta}
& = & \varepsilon^{\alpha/\beta}_{\rm DFT}-\varepsilon^{\alpha/\beta}_{\rm GW} \, .
\end{eqnarray}
Here $\varepsilon^{\alpha}_{\rm DFT}$ ($\varepsilon^{\beta}_{\rm DFT}$) identifies the
QP-energy prediction at a given high-symmetry point of the conduction band (valence band) as described in  PBE, CX, KI-PBE, or KI-CX, while 
$\varepsilon^{\alpha/\beta}_{\rm GW}$ represents the corresponding reference value. The gray horizontal dashed line (at zero deviation) permits an easy comparison among different types of QP gaps.
Color shading highlights the deviation that corresponds to the fundamental QP gap, as consistently determined in KI-DFT and $GW$ for the AA$'$ system.

Figure \ref{fig:error_all}
shows that use of a KS-DFT
approach (in CX and PBE)
persistently leads to an underestimation of the gaps for QP transitions. 
Use of the corresponding KI-DFT forms leads instead to small overestimation of gaps and significant reductions in the absolute deviation values. That is, PBE and CX are both good starting points for a KI-DFT characterization of QPs overall, when investigating structures fixed at their experimental geometry. 

We observe that, for the transitions $\text{G}_{v}\rightarrow \text{G}_{c}$, $\text{K}_{v}\rightarrow \text{G}_{c}$, and $\text{M}_{v}\rightarrow \text{G}_{c}$, the KI-CX predictions
are closer to $GW$, while for the rest of the transitions, KI-PBE are slightly closer to 
$GW$. Both methods for KI-DFT give
predictions for band gaps (originating at the set of relevant conduction- and valence-band extrema) that remain within 1 eV from corresponding $GW$ results, taking the form of a near-constant offset. The latter observation suggests that both KI-PBE and KI-CX give an accurate treatment of transitions involving the conduction band $\text{G}_{c}$ symmetry point
for bulk systems.

\subsection{Fingerprints of vdW: QP bandstructures at CX- and PBE-predicted atomic structures\label{sec:fingerprints}}

\begin{figure}
    \centering
    \includegraphics[width=1\linewidth]{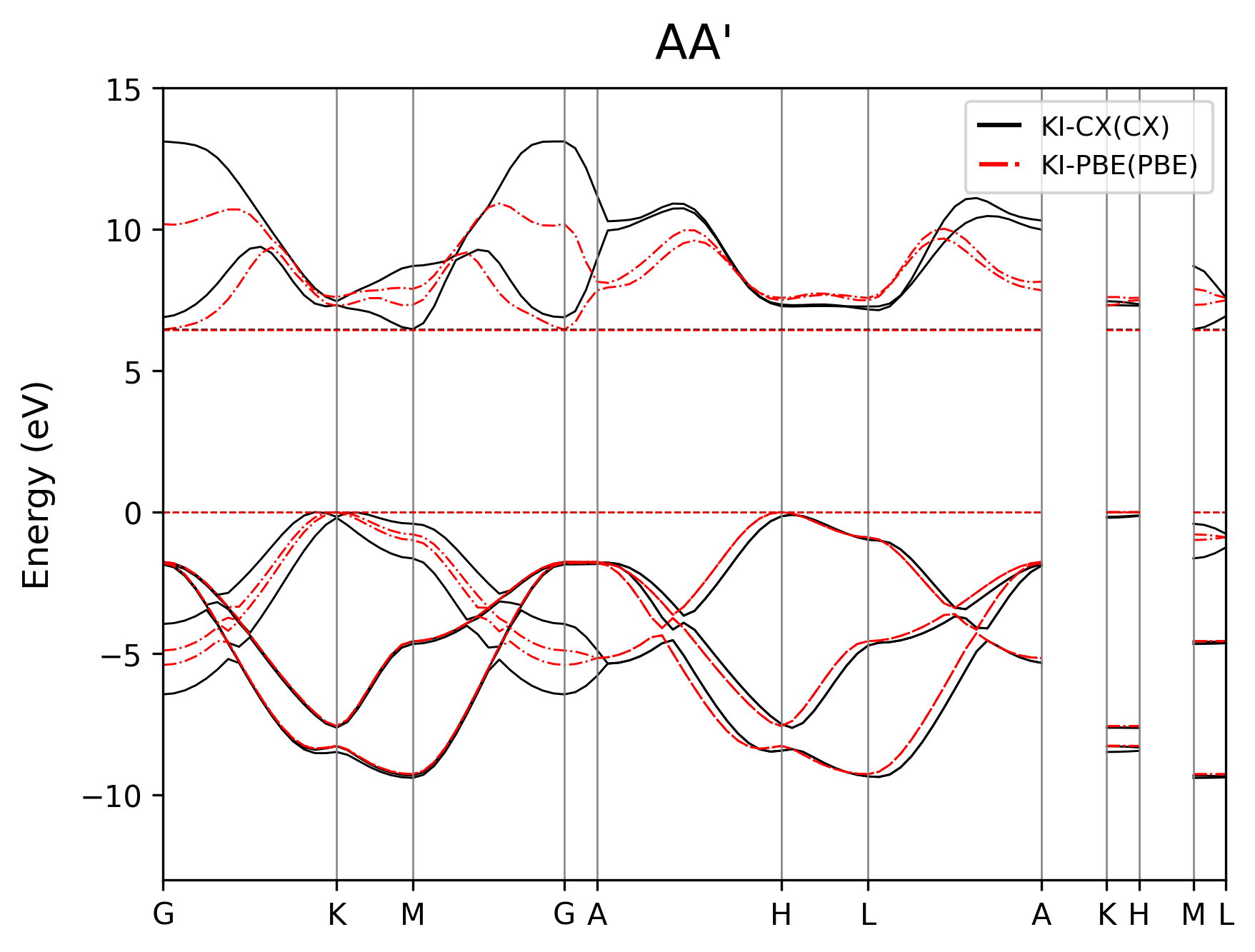}
    \caption{Comparison of QP predicted for AA$'$
    by KI-CX(CX) and KI-PBE(PBE). Here the KI-CX(CX) QP predictor 
    results when we use KI-CX at the atomic configuration predicted by a CX-based KS-DFT structure optimization; KI-PBE(PBE) results instead when we systematically stick to PBE in the modeling procedure. As the atomic structures differs, we shift the pair of resulting KI-DFT bandstructures differently, i.e., by their respective (functional-specific) 
    predictions for the VBM-energy values.}
    \label{fig:aap_bands}
\end{figure}

We finally provide analysis that substantiates our interpretation of all so-far discussed major QP changes as indirect vdW-type (or nonlocal-correlation) signatures. Our discussion leads us to identify additional such signatures on the here-predicted QP nature for 
bulk h-BN. 

We first define general vdW fingerprints on the QP bandstructure and recall the logic of sorting these into what we call direct and indirect features, for example, as discussed in 
Ref.\  \onlinecite{rohrer11p165423}. We note that proximity of interacting fragments (in our case h-BN layers) impacts the nature \cite{signatures,hyldgaard2020} as well as the scaling and strength of vdW forces \cite{kleis08p205422,NanovdWScale}. We
expect stronger vdW signatures to appear also
in the QPs bandstructure \cite{rohrer11p165423} as the interlayer separation shortens, something that the inclusions of the vdW-DF $E_c^{\rm nl}$ term itself drives. Use of CX in KI-DFT as well as in a KS-DFT structure optimization means that we 1) include vdW attraction between the layers and thus correct for the large (interlayer) underbinding that exists in a GGA like PBE \cite{rydberg03p606,rydberg2003,shukla2022} and, in turn, that we 2) provide a \textit{consistent-KI-CX} QP characterization, covering vdW impacts to the extent that they are included in CX. The correspondingly defined `KI-PBE(PBE)' approach should only be used as a reference for discussions when vdW forces are important.

By contrasting results for AA$'$ by KI-CX(CX) and 
KI-PBE(PBE), below, with those of the fixed-structure KI-CX and KI-PBE method, above, we arrive at a precise definition of direct and indirect vdW fingerprints in QP predictions. 
KI-CX(CX) versus KI-PBE(PBE) differences are nominally a full description of vdW and other truly nonlocal-correlation fingerprints on the QP nature. The same holds in practice for differences between KI-CX(CX) results for AA$'$ and KI-CX characterizations of BN1 (because a switch from PBE to CX has little effect on the description of intralayer bonding). Meanwhile, KI-CX versus KI-PBE differences are denoted as  `direct' vdW band-structure impacts 
because they reflect the same atom configuration and track the
impact of introducing the truly nonlocal correlation term $E_c^{\rm}$ for what is, in practice, nearly identical electron-density variations $\rho(\mathbf{r})$; The similarity 
of CX and PBE predictions for $\rho(\mathbf{r})$ at a fixed structure
has been validated by comparing workfunction shifts and vibrations 
across many types of systems \cite{hyldgaard2020,grwahy20,frostenson2022}. 
We identify `indirect
vdW signatures' on our QP predictions, when
KI-CX(CX) versus KI-PBE(PBE) differences
cannot be identified as a direct vdW fingerprint. 

Before turning to specific interpretations, we
express a general expectation that indirect vdW effects on QPs will be dominant for layered systems \cite{rohrer11p165423}. This follows because PBE and hence KI-PBE(PBE) cannot drive a layered bulk system to any meaningful interlayer proximity as discussed both previously \cite{rydberg03p606,rydberg2003}
and as computed here:  It is alone our CX prediction of structure that is close to the experimental characterizations of h-BN, see Section III. In any case, having the capability, via KI-CX(CX) studies, to predict both indirect
and full vdW effects, is essential when we seek to predict QPs either independently from an existing experimental structure characterization \cite{RanPRB16}, or in the absence of sufficient structure data \cite{rohrer11p165423,frostenson2022,frostenson2024,frostenson2024b}, or even ahead of actual material synthesis \cite{Sofo2007,rohrer11p165423}. 
 
Figure~\ref{fig:aap_bands}
reports our comparison of the QP bandstructure predicted in KI-CX(CX), solid black line, and in KI-PBE(PBE), red dashed-dotted line. As stated in the previous sections, for these QP predictions we now  employ what we can call native interlayer distances. That is, we set $c$ lattice constant for the AA' system
by a CX-based structure optimization (giving $c=6.448$ {\AA}) for the ensuing KI-CX study and a PBE-based  optimization (giving $c=8.560$ {\AA})  
for the KI-PBE(PBE) study.

Contrasting the KI-CX(CX) and KI-PBE(PBE) QP results, we first focus on the valence bands along the symmetry path $\text{G}_{v} \to \text{K}_{v} \to \text{M}_{v} \to \text{G}_{v}$. At a general level, we find that use of KI-CX(CX) as our QP predictor lifts the near degeneracy found in KI-PBE(PBE) for the two top-most valence bands around the `K', placing the VBM at the non-symmetric $k$-point $\text{T}_{v}$. 
We observe that this KI-CX(CX) VBM 
shift, $\text{K}_{v} \to \text{T}_{v}$, is in qualitative agreement with both KI-CX and the KI-PBE predictions obtained at the experimental value for the $c$ lattice constant, Fig.\ \ref{fig:main_bands}, lower-right panel. That is, the change is primarily a consequence of structure and should be considered as an indirect effect of including vdW interactions. We add that this VBM shift is also found when dispersion-corrected HSE-D3 \cite{HSE03,HSE06,grimmeD3} is used to predict
the band structure of the AA$'$ structure \cite{wickramaratne2018}.

The shift to $\text{T}_v$ of the QP VBM for 
the AA$'$ structure (bottom section of Table \ref{tab:quasiparticle_gaps}) is caused by a general type of nonlocal-correlation (or beyond-pure-vdW) effect that we call incipient covalency \cite{langreth2009,Berland_2015:van_waals}. This VBM shift also arises for the AB structure,  middle section of Table \ref{tab:quasiparticle_gaps} and exists in bulk h-BN because the VBM orbital have a pronounced $p_z$ character \cite{wickramaratne2018}, see Fig.\ \ref{fig:conduct_band_minimum}. The fact that these occupied orbitals extend into the inter-layer regions means on the one hand that they enhance the vdW attraction \cite{hyldgaard2020} and on the other hand that such nonlocal-correlation effects drive a reduction of the interlayer separation (to close to the experimental $c$ lattice constant of bulk h-BN) compared with descriptions by semilocal functionals \cite{rydberg03p606,rydberg2003}, 
e.g. PBE. The overall implication is that the DFT inclusion of the nonlocal-correlation term $E_c^{\rm nl}$ (sometimes equated as simply vdW forces) also -- and in fact mainly -- causes indirect QP signatures \cite{rohrer11p165423}, 
for example, a small hybridization of slightly overlapping $p_z$ orbitals originating from different layers.

In the upper hexagon of high-symmetry $k$-points in the BZ (corresponding to the A-H-L-A path), we note that use of KI-CX(CX) as our QP predictor generally causes only small shifts of KI-PBE-(PBE) predictions for the occupied-state energies; These shifts are upwards in some and downwards in other parts of this region of the BZ. However, in the paths of $k$ points that connects the upper and lower parts of the irreducible BZ, 
we find that use of KI-CX(CX) causes QP-dispersion changes in the 
$\text{G}_{v}$ - $\text{A}_{v}$ and 
$\text{M}_{v}$ - $\text{L}_{v}$ variations. 

The major differences between KI-CX(CX) and KI-PBE(PBE) predictions arise for 
the QP conduction bands,
Fig.\ \ref{fig:aap_bands}.
For example, KI-CX(CX) predicts a $\sim 3$ eV larger dispersion 
in the BZ region near G
(for the lowest-lying conduction bands).
We find that KI-CX(CX) predicts a CBM at the $\text{M}_{c}$ symmetry point, while the KI-PBE(PBE) predictor places it near G. 
The difference arises because 
KI-CX(CX) can be seen as raising the conduction-band QP values at the $\text{G}_{c}$ point, making KI-CX(CX) consistent with $GW$ on the CBM \cite{hunt2020}. We can interpret these effects as again being mostly indirect vdW signatures. Nevertheless,
even when we compared the KI-CX and KI-PBE at the same (experimental) structure, we did find some
differences for the energy-variation in QP conduction-band 
predictions.

Finally, we can interpret the 
accuracy of the KI-CX(CX) predictor in terms of 
vdW interactions producing a lowering of the $\text{T}_{v}\rightarrow \text{M}_{c}$ QP energy ($6.47$ eV). This is consistent with the observation that a closer 
(vdW-induced) proximity of the layers causes an overlap of the $p_z$ orbitals, something that KI-PBE(PBE)
entirely misses. Again, the absence of a consistent vdW and general nonlocal-correlation inclusion in KI-PBE(PBE) causes it to predicts a $\text{K}_{v}\rightarrow \text{G}_{c}$ transition as the fundamental gap, at $7.32$ eV.

\subsection{Broader relevance of KI-CX usage}

We observe that moving towards a full and consistent inclusion of vdW interactions in the descriptions of both QPs and GS structure and energies (the traditional role of DFT) is no easy task.
Traditional MBPT 
methods, like GW, and wavefunction methods,
like EOMCC, are today practical computational
tools that can be used to characterize excited-state properties. However, these are costly compared to KI-DFT
(that has the cost of DFPT) or gKS methods (that have the
cost of hybrids). Also, a full capture of vdW-interaction effects requires self-consistent GW with vertex corrections (GW$\Gamma$)  \cite{kutepov2021,weng2024}. Similarly, wave-function methods can predict charge-excited states with high accuracy, for example 
EOMCC for ionized states \cite{stanton1994,barlett1997,quintero2023} or EOMCC for electron-attached states \cite{nooijen1995}. However, in such a framework, it is expected that the use
of symmetry-adapted perturbation theory is necessary for incorporation of vdW effects \cite{patkowski2020,szalewicz1979}. 

We consider KI-CX (which reduces to use of CX in KS-DFT when used for structure optimizations) as a faster alternative to GW and EOMCC studies when seeking to predict QPs and identify fingerprints of vdW interactions \cite{RanPRB16}. The CX functional is, as a conserving vdW-DF \cite{dion2004,berland2014,hybesc14,hyldgaard2020}, designed to capture the vdW interactions directly within the MBPT framework, as far as possible. Of course, it may not capture all nonlocal-correlation effects and that is why continued testing (e.g., via KI-CX predictions) is important.

We repeat our observation that KI-CX(CX) stands out by proceeding with the atomic-structure and QP characterizations in concert. As stated in the introduction, the CX (in KS-DFT) and KI-CX (in KI-DFT)
use exactly the 
same MBPT input to define the nonempirical XC functional, per se. We think that a possible success of the CX design as a dual-use (KS-DFT and KI-DFT) functional, is helped by its inherent emphasis on current conservation and thus, the implied reflection of the plasmon nature. On the one hand, the plasmon-energy shifts by electro-dynamical coupling \cite{Berland_2015:van_waals,hyldgaard2020} is a natural framework to count the integrated effects of
nonlocal-correlation effects \cite{jerry65,hybesc14}. On the other hand, a consistent modeling of
virtual currents (as in plasmons) are  
essential when moving electrons (as in the
excitations that QPs characterize).

Our interest in KI-CX goes well beyond the present focus on testing. By use of KI-CX we may be able to provide fast predictions of atomic and electronic (QP) structure in concert, for example, tracking fingerprints of vdW and, in general, noncovalent interactions in DNA systems while comparing with, e.g., optimally-tuned RSH vdW-DFs  \cite{JPCM2025}.

\section{Summary and outlook}

A comparative analysis of results reveals systematic trends across h-BN systems. The data is obtained by PBE and CX functions in KS-DFT functionals (denoted PBE, CX results) and with the corresponding use in KI-DFT (giving corresponding KI-PBE, KI-CX results). The performance is compared against $GW$ results, where available. For both PBE and CX, we find that their use in KI-DFT consistently leads to better performance on QP descriptions 
that what their traditional (and originally intended) usage gives in KS-DFT.

For our present interest in systems with vdW and noncovalent binding, the
KI-CX holds advantages over the traditional KI-PBE. We have considered both direct and indirect effects of the nonlocal-correlation effects on the QP. The former (latter) type of nonlocal-correlation effects are revealed when considering the QP band-structure predictions
by KI-CX and KI-PBE at the same structure 
(when we consider the
problem at the optimal
atomic structure defined 
by KI-CX and KI-PBE).
We find that KI-CX 
is significantly better that KI-PBE at QP predictions when we 
also use these functional to first predict the atomic structure.

Furthermore, we find that KI-CX delivers accuracy at QP predictions also when asserted on just the direct effects of including the vdW-type effects reflected in $E_c^{\rm nl}$ (i.e., when we compare KI-CX and KI-PBE at the same structure, fixed by experimental observations). For a summary, we focus on the 
quality at predicting the 
set of bulk-hBN QP gaps,  i.e., energy differences between local QP-valence-band minima and local QP-conduction band maxima at various high-symmetry points of the Brillouin zone, e.g., $\text{G}_{c}$, $\text{K}_{c}$.
While all of the investigated methods (use of CX or PBE in KS-DFT or as KI-CX and KI-PBE) exhibit transition-dependent errors (as asserted by comparison with GW), the KI-CX remains generally close to the GW values for the gaps. 
Overall, we find that the the KI-CX results for the QP band structure validates the CX as a robust XC, performing also here at least as well  as PBE.

We intend to use KI-CX to seek an acceleration of predictions of (ionizing) optical response of
molecules, semi-conductors and insulators.
An advantage of using what we call consistent-KI-CX, or KI-CX(CX), lies in the fact that it is set up to predict QP in complex materials (cases when there is only a partial experimental
characterization of atomic structure details).
In those cases we need
a robust, general-purpose vdW-DF, like CX, to first establish the details of atomic structure in KS-DFT. With the KI-CX(CX) method, we argue that one can immediately extract predictions for QP properties (simply as KI-DFT states instead of KS-DFT states). This can, in principle, be done ahead of actual synthesis and ensuing structure predictions for general materials that are impacted by vdW forces.

\begin{acknowledgments}
We thank Nicola Colonna and Edward Linscott for valuable discussions.
The present work is supported by the Swedish Research Council (VR) through Grant No.\ 2022-03277, 
and by the Chalmers Area of Advance (AoA) Nano
and Chalmers AoA Production.
The computations were performed using computational and storage resources at 
Chalmers Centre for Computational Science and Engineering (C3SE), 
and with computer and storage allocations from the 
National Academic Infrastructure for Supercomputing in Sweden (NAISS), under contracts
NAISS2023/3-22, 
NAISS2023/6-306, 
NAISS2024/3-16, 
and NAISS2024/6-432. 

\end{acknowledgments}

\begin{table*}
    \centering
    \begin{tabular}{l c | c c c c c c}
    \hline
    $c$ [{\AA}] & $E_{\rm cut}^{\rm wf}$ [Ry] & $\text{G}_{v}\rightarrow \text{G}_{c}$ & $\text{M}_{v}\rightarrow \text{M}_{c}$ & $\text{K}_{v}\rightarrow \text{G}_{c}$ & $\text{K}_{v}\rightarrow \text{K}_{c}$ & $\text{M}_{v}\rightarrow \text{G}_{c}$ & $\text{K}_{v}\rightarrow \text{M}_{c}$ \\
    \hline
    20 & 100  & 7.62 & 8.98 & 5.90 & 7.94 & 6.79 & 8.08 \\
    20 & 200  & 8.14 & 9.90 & 6.42 & 8.61 & 7.31 & 9.00 \\
    20 & 300 & 8.18 & 9.86 & 6.46 & 8.62 & 7.36 & 8.96 \\
    \textit{20} & \textit{400}  & \textit{8.21} & \textit{9.92} & \textit{6.48} & \textit{8.69} & \textit{7.38} & \textit{9.02} \\
    \hline
    30 &100 & 8.36 & 10.01 & 6.63 & 8.81 & 7.53 & 9.11  \\
    30 &200 & 8.28 & 9.95 & 6.55 & 8.65 & 7.45 & 9.05 \\
    30 &300 & 8.28 & 10.17 & 6.56 & 8.89 & 7.45 & 9.27 \\
    30 &400 & 8.20 & 9.99 & 6.47 & 8.72 & 7.37 & 9.09 \\
        \hline    
    \end{tabular}
    \caption{Search for precision (numerical stability with respect to choice of modeling parameters) in KI-CX predictions of BN1 QP-gap energies (in eV) as defined by transitions between indicated valence- and conduction-band symmetry points. Italicized table entries 
    identify the set of model parameters (along with corresponding gap results) that we use, in the main text, to report and discuss the QP behavior for the strictly two-dimensional BN1 system.}
    \label{tab:bn1Converge}
\end{table*}

\appendix

\section{Linear response for nonlocal correlations}

As mentioned in Section~II.B, the DFPT response of the non-local potential $v^{\text{nl}}_{}(\mathbf{r})$ is implemented in the \textsc{kcw} module of QE. The necessary linear-response steps are derived in Ref.\  \onlinecite{sabatini2016} with respect to a general perturbation $\Delta^{\lambda} v^{\text{nl}}_{}(\mathbf{r})$ and with the purpose (there) of computing vdW signatures on phonons and their interactions. Here, we summarize how this DFPT formulation for $\Delta^{\lambda} v^{\text{nl}}_{}(\mathbf{r})$
appears when considered with respect to the removal or addition of a fractional electron, i.e., for KI-DFT use.

As a starting point, consider the definition of the non-local energy in Eq.~\eqref{eq:nlenergybykernel}, approximating the spatial integrals by sums over a uniform grid of points (at fixed positions $\rho_i$ and $\mathbf{r}_i$) and the gradients by finite differences on the same grid. This allows the use of partial derivatives of the density variation on the grid to be used
in place of formal functional derivatives \cite{roso09}. Applying these finite differences and simplifying, we obtain:
\begin{equation}
   E^{\text{nl}}_{c} = \sum_{\alpha} \sum_{i} u_{\alpha i}(\mathbf{r}_i) \Theta_{\alpha i}(\mathbf{r}_i),
   \label{eq:grid_nl_e}
\end{equation}
where the first factor corresponds to the auxiliary function
\[
u_{\alpha i} = \sum_{\beta} \sum_{j} \phi_{\alpha \beta}(\mathbf{r}_{ij}) \Theta_{\beta j}(\mathbf{r}_j).
\]
Here, $\Theta_{\alpha i} = \rho_i P_{\alpha}[q(\rho_i, |\nabla \rho_i|)]$, the values $q_{\alpha}$ are chosen to ensure accurate interpolation of the function $\phi$, and $P_{\alpha}$ is the function resulting from the interpolation.

From Eq.~\eqref{eq:grid_nl_e}, one can obtain the non-local potential ($\partial E^{\text{nl}}_{c} / \partial \rho$) in a more compact notation:
\begin{eqnarray}
    v^{\text{nl}}_{ i} & = & \sum_{\alpha} \left[ \qty(u_{\alpha i} \frac{\partial \Theta_{\alpha i}}{\partial \rho_i}) + \sum_{j} u_{\alpha j} \qty(\frac{\partial \Theta_{\alpha j}}{\partial \abs{\nabla \rho_{j}}} \frac{\partial \abs{\nabla \rho_{j}}}{\partial \rho_{i}}) \right] 
    \nonumber \\
    & \equiv & b_{i} + \sum_{j} h_{j} \, ,
    \label{eq:compact_vnl}
\end{eqnarray}
where $b_i$ corresponds a term defined exclusively
by differentiating direct by density (at the specific grid site $i$). The second term, defined by a set of
$h_j$ contributions, reflects instead derivations with respect to the density gradients. The evaluation of a gradient in QE makes use of specific stencil so that the gradient 
at site $i$ can be expressed in terms of the density values at neighboring grid sites $j$ \cite{roso09}.

To obtain the non-local response kernel for $E_{\rm c}^{\rm nl}$, Eq.~\eqref{eq:nl-lrkernel}, it is necessary to apply DFPT to Eq.~\eqref{eq:nlenergybykernel}. To do this, consider an infinitesimal fraction of an electron $\delta\mathbf{0} n$ added or removed to the $n$th MLWF at $\mathbf{0}$.  This generates a perturbation in the effective KS 
potential $\Delta^{\delta\mathbf{0} n} v_{\text{KS}}$, leading, in turn to a new variation of the density, with relative changes denoted $\Delta^{\delta\mathbf{0}n} \rho(\mathbf{r})$. 
The first-order change in this non-local potential is:
\begin{equation}
    \Delta^{\delta\mathbf{0} n} v^{\text{nl}}_{i} = \sum_{j} \frac{\partial v^{\text{nl}}_{i}}{\partial \rho_{j}} \Delta^{\delta\mathbf{0} n} \rho_{j} \, .
    \label{eq:finit_nl}
\end{equation}

Expressing also the  $v^{\text{nl}}_{i}$
for in Ref.\ (\ref{eq:finit_nl}) reveals the second derivatives present in Eq.~\eqref{eq:nl-lrkernel}. To get a numerical evaluation, we adapt the grid-specific implementation \cite{roso09,sabatini2016,giannozzi2017} of the formal determination 
\begin{equation}
    \sum_{ij} \frac{\partial^{2} E_{c}^{\text{nl}}[\rho_i]}{\partial \rho_{i} \partial \rho_{j}} \Delta^{\delta\mathbf{0} n} \rho_{j} \approx \iint d\mathbf{r}  d\mathbf{r}'  \frac{\delta^{2} E^{\text{nl}}_{c}}{\delta \rho(\mathbf{r}) \delta \rho(\mathbf{r}')} \Delta^{\delta\mathbf{0} n} \rho(\mathbf{r}') \, .
\end{equation}

Finally, to get the DFPT, we can simply linearize the two components of Eq.~\eqref{eq:finit_nl}:
\begin{equation}
    \Delta^{\delta\mathbf{0} n} v^{nl}= \sum_{i} \qty(
    \frac{\partial b_{j}}{\partial \rho_{i}}+
    \sum_{k}\frac{\partial b_{k}}{\partial \rho_{i}}
    )\Delta^{\delta\mathbf{0} n} \rho_{i} \, .
    \label{eq:Deltafinit_nl}
\end{equation}
The details of these two components are explicitly derived, for the stencil used in QE, in Ref.\ \onlinecite{sabatini2016}. 

In summary, KI-DFT in QE is in fact already set up to 
move to a KI-CX(CX) form 
and general-purpose starting point for QP characterizations, as here illustrated.

\section{Precision in KI-CX}

For KS-DFT studies using norm-conserving pseudopotentials, past experience suggest that a choice $E_{\rm cut}^{\rm wf}=200$ Ry for wavefunction kinetic-energy cut off is sufficient to converge 
predictions of total-energy differences and  structure \cite{shukla2022,JPCM2025}. 

We find that the same choice also consistently leads to robust predictions for the valence-type QP energies. This is expected and can be understood from the following observations.
First, the KS-DFT studies can only converge the total energy and the density variation when the spatial variation in the occupied KS-DFT orbitals are all themselves converged. Second, we find vanishing differences, besides 
an overall energy shifts, between the valence
band structure as described in, for example, CX and KI-CX. Our experience is that the process of making the MLWF representations, and hence the completion of the KI-CX functional specification, is inherently robust for descriptions of the set of occupied QPs states.

For descriptions of conduction-band QP states we see a
need to seek numerical stability with regards to changes in
choices for implicit modeling parameters; This observation has 
practical implications for our KI-CX study of the 
single-sheet BN1 system. The description of unoccupied QPs is 
generally a challenge for KS-DFT, for example, documented for 
electron attachment in nitrogen-base systems \cite{nguyen2016,JPCM2025}. 
This problem emerges, in part, because the spatial variation of 
unoccupied KS-orbitals does not naturally enter in the iterative 
optimization of the actual ground-state electron density
$\rho(\mathbf{r})$. In contrast, with a hybrid like AHBR-mRSH and 
with the KI-DFT framework, we can overcome also this KS-DFT problem 
to accurately describe the empty QPs \cite{nguyen2016,JPCM2025} 
because we make the iterative computational scheme dependent on the
nature of these empty orbitals. However, the requirements on the 
$E_{\rm cut}^{\rm wf}$ choice for stability in the description of 
conduction-band states might well be higher. Also, given our screening 
handling \cite{gygi1986}, we must validate that the chosen height $c$ of the 
modeling cell for the strictly two-dimensional
BN1 system is sufficient to electrically decouple the sheets.

Table \ref{tab:bn1Converge} presents an overview of the 
numerical work we completed to assert the level of precision we 
presentlty have for a KI-DFT description of (empty-state) QP energies
in BN1. We see that the step of crafting the MLWF representation can 
sometimes cause abrupt changes in the resulting predictions of 
conduction-band QP energies.  A complication is that the 
$E_{\rm cut}^{\rm wf}$ variation also depends on the choice of $c$. 
Our criterion for using these BN1 QP predictions is that we reach 
a fair numerical stability with regards to changes in both these 
modeling parameters. Tables \ref{tab:bn1} and 
\ref{tab:bn1Converge} report our finding
that we have a fair cross-checked stability 
for the conduction-band QPs of BN1 when using
$c=20$ {\AA} and $E_{\rm cut}^{\rm wf}=400$ Ry, as also stated in the main text.

We also performed a study of numerical stability
for the bulk AA$'$ system in KI-CX but found no variation in 
predictions for QP gap results as we increased $E_{\rm cut}^{\rm wf}$ 
from 150 Ry to 300 Ry. Accordingly we choose to complete all 
KI-DFT characterizations for bulk h-BN at the value 
$E_{\rm cut}^{\rm wf}= 200$ Ry that is adequate for total-energy 
and structure characterizations in KS-DFT.

\providecommand{\noopsort}[1]{}\providecommand{\singleletter}[1]{#1}%

\end{document}